\documentclass[pra,aps,twocolumn,floatfix,superscriptaddress]{revtex4-2}
\usepackage{amsmath}
\usepackage{amssymb}
\usepackage{graphicx}
\usepackage{color}
\usepackage{soul}
\usepackage[utf8]{inputenc}
\usepackage[T1]{fontenc}
\usepackage{array}
\usepackage{comment}
\usepackage{amsmath}
\usepackage{mathtools}
\newcommand{\abs}[1]{\left\vert #1 \right\vert } 
\usepackage{braket}

\usepackage[normalem]{ulem}

\begin{document}

\title{Spectrum and quench-induced dynamics of spin-orbit coupled quantum droplets}

\author{Sonali Gangwar}
\affiliation{Department of Physics, Indian Institute of Technology, Guwahati 781039, Assam, India} 

\author{Rajamanickam Ravisankar}
\affiliation{Department of Physics, Zhejiang Normal University, Jinhua 321004, PR China}
\affiliation{Zhejiang Institute of Photoelectronics $\&$ Zhejiang Institute for Advanced Light Source, Zhejiang Normal University, Jinhua, Zhejiang 321004, China}

\author{S. I. Mistakidis}
\affiliation{ITAMP, Center for Astrophysics, Harvard $\&$ Smithsonian, Cambridge, MA 02138 USA}
\affiliation{Department of Physics, Harvard University, Cambridge, Massachusetts 02138, USA}

\author{Paulsamy Muruganandam}
\affiliation{Department of Physics, Bharathidasan University, Tiruchirappalli 620024, Tamilnadu, India}

\author{Pankaj Kumar Mishra}
\affiliation{Department of Physics, Indian Institute of Technology, Guwahati 781039, Assam, India}

\date{\today}

\begin{abstract} 

We investigate the ground state and dynamics of one-dimensional spin-orbit coupled (SOC) quantum droplets within the extended Gross-Pitaevskii approach. As the SOC wavenumber increases, stripe droplet patterns emerge, with a flat-top background, for larger particle numbers. The surface energy decays following a power-law with respect to the interactions. At small SOC wavenumbers, a transition from Gaussian to flat-top droplets occurs for either a larger number of atoms or reduced intercomponent attraction. The excitation spectrum shows that droplets {for relatively small SOC wavenumbers are stable, otherwise stripe droplets feature  instabilities as a function of the particle number or the interactions.} 
We also witness rich droplet dynamical features using velocity imprinting and abrupt changes in the intercomponent interaction or the SOC parameters. Characteristic responses include breathing oscillations, expansion, symmetric and asymmetric droplet fragmentation, admixtures of single and stripe droplet branches, and erratic spatial distributions suggesting the triggering of relevant instabilities. Our results reveal the controlled dynamical generation and stability {properties} of stripe droplets that should be detectable in current cold-atom experiments.

\end{abstract}

\flushbottom

\maketitle
\section{Introduction}

Cold atoms constitute fertile platforms to unravel a diversity of many-body phenomena, such as supersolidity~\cite{chomaz2022dipolar}, quasiparticles~\cite{massignan2014polarons, mistakidis2023few}, Anderson localization~\cite{modugno2010anderson, Sarkar2023}, and self-bound states of matter e.g. quantum droplets QDs)~\cite{Petrov2015, Luo2020}. The latter appears as the result of the competition between repulsive mean-field interactions and attractive quantum fluctuations in one-dimension (1D) being the focus of our study. 
The lowest-order quantum correction is represented by the Lee-Huang-Yang (LHY) term~\cite{Lee1957}, which in turn leads to the concept of the extended Gross-Pitaevskii equation (eGPE)~\cite{Petrov2015, Petrov2016, Luo2020} describing self-bound states; see also Refs.~\cite{Parisi2019, Mistakidis2021, ota2020beyond, Englezos2023} for beyond LHY effects. 
Historically, QDs have been first experimentally observed in dipolar gases~\cite{FerrierBarbut2016, Schmitt2016, bottcher2020new} and much later in mixtures thereof~\cite{trautmann2018dipolar, politi2022interspecies}. Meanwhile, they were realized in binary short-range homonuclear~\cite{Cheiney2018, Semeghini2018} and heteronuclear~\cite{DErrico2019, Burchianti2020} bosonic settings.

In this latter case, which we also explore herein, the main focus of research has been to examine the ground state droplet configurations with respect to the involved atom number and mean-field interactions~\cite{Luo2020}. For instance, it was explicated that a decreasing intercomponent attraction or larger particle number leads to a transition from Gaussian (or soliton-like) droplet distributions to flat-top (FT) ones~\cite{Petrov2015, Luo2020, mistakidis2023few}. Specifically, when FT is reached by loading more atoms renders the configuration wider while maintaining the same peak density manifesting droplet incompressibility. Another interesting feature of droplets is their well-defined surface tension, which, together with their kinetic energy, plays a crucial role in achieving their stability when the former dominates~\cite{Astrakharchik2018, Luo2020, Ferioli2019}. Moreover, basic properties of their underlying excitation spectrum~\cite{Tylutki2020, katsimiga2023interactions}, triggering of modulational instability events~\cite{Mithun2020, mithun2021statistical} and the impact of nonlinear excitations such as dark solitons~\cite{Edmonds_dark_drops, katsimiga2023solitary} and vortices~\cite{li2018two, kartashov2022spinor} were discussed. Very recently, other generically unstable bound states called bubbles~\cite{Edmonds_dark_drops, katsimiga2023interactions} existing at more negative chemical potentials have been identified. On the other hand, notable examples of the droplet dynamical response include their collision properties in one-~\cite{Astrakharchik2018, katsimiga2023interactions}, two-~\cite{hu2022collisional} and three-dimensions~\cite{Ferioli2019}. It was showcased that mainly slow-moving droplets merge, and fast ones feature quasi-elastic collisions. Especially in 1D, they can even fragment~\cite{Astrakharchik2018}, while their interactions can be explained through an effective particle picture~\cite{katsimiga2023interactions}. Additionally, scattering processes of 1D droplets against a potential well in terms of their velocity and particle number were analyzed~\cite{debnath2023interaction}. 

The majority of the above investigations were based on the reduction of the genuine two-component mixture into an effective single-component one (justified under specific symmetry conditions~\cite{Petrov2015, Luo2020}) but also in the absence of spin-orbit coupling (SOC). 
To appreciate the role of intercomponent interactions, the first steps towards the two-component system have been taken recently without~\cite{mistakidis2021formation, valles2023quantum} and with SOC~\cite{Tononi2019, Gangwar2022, Gangwar2023, Li2017a, cui2018spin}. An immediate additional feature that SOC brings into play is the formation of stripe {patterns~\cite{hou2018momentum, qu2013observation,Ravisankar2021b}}, which here build upon the droplet background as it was shown for varying mean-field interactions in~\cite{Gangwar2023}. {In other direction Sachdeva et al. demonstrated the transition from the supersolid stripe phase to the zero-momentum droplet upon increasing the Rabi coupling for fixes SOC in two dimension~\cite{PhysRevA.102.043304}.
This characteristic poses fundamental questions related to the underlying excitation spectrum that remains elusive and could allow us to infer the structural stability of SOC droplets. 
In this context, also alterations between phononic and rotonic-like excitations could be revelead.} Along the same lines, the behaviour of the surface energy of the stripe droplet is not yet understood. 
Another intriguing direction concerns the dynamical response of these droplet configurations. For vanishing mean-field interactions, where soliton structures take place~\cite{Gangwar2022}, it was argued that their dynamics, following velocity perturbations or interaction quenches, exhibit breathing oscillations, moving states, spin flipping or generation of secondary solitons. For finite interactions, the droplet was shown to be more stable against perturbations~\cite{Gangwar2023}. 
It is, however, still open whether, for finite interactions, other response regimes can be entered, such as droplet fragmentation or spontaneous generation of stripe droplets due to the interplay of different energy contributions. In the present work, we tackle these issues by employing a 1D symmetric SOC bosonic mixture, which one could describe within a corresponding two-component eGPE model. 

We find that for large SOC wavenumbers and finite interactions, stripe QDs occur, and their density background saturates to an FT for increasing atom number. On the other hand, at small wavenumbers, it is possible to tune the transition from Gaussian to FT droplet distributions for larger (smaller) atom numbers (due to intercomponent attraction) \cite{Astrakharchik2018}. The kinetic energy shows an interaction-dependent maximum at the transition threshold, while the surface energy decays in a power-law fashion as a function of the relative interaction ratio for both stripe and standard droplets. The stability of the configurations is studied through their excitation spectrum, constituting a central result of our work. 
{In the case of small SOC wavenumbers, Gaussian and FT droplets are stable irrespective of the atom number or the magnitude of the interactions. This result is in line with the spectral stability of symmetric droplets appearing in short-range interacting bosonic mixtures~\cite{Tylutki2020, katsimiga2023interactions}. However, further increasing the SOC wavenumber where stripe droplets with FT background occur instabilities arise with respect to both the atom number and the intercomponent attraction.}

The dynamical response of the SOC droplets is dictated by the interplay of the involved energy contributions. We mainly consider four distinct situations: (i) initial velocity and quenches on (ii) the intercomponent attractive coupling, (iii) the Rabi coupling, and (iv) the SOC wavenumber. Focusing on velocity perturbations, we exemplify that irrespective of the SOC wavenumber and for increasing velocity, the droplet either performs breathing motion or breaks into moving droplets since the attractive SOC energy term prevails. The interaction-dependent character of the breathing frequency and critical velocity (moving droplets) is analyzed. Specifically, the breathing frequency is smaller for stripe droplets, and the FT region features a power-law decay with the inverse of the particle number. Turning to intercomponent interaction quenches, we show that for larger quench amplitudes: (i) reducing the attraction leads either to droplet expansion or splitting into two counterpropagating fragments, and (ii) increasing the attraction gives rise to a droplet breathing or breaking into several fragments. The sudden modifications on the Rabi-coupling are associated with the generation of a stripe droplet and asymmetric droplet fragmentation processes. Finally, quenches on the SOC wavenumber enforce the nucleation of both single and stripe droplet branches or erratic spatial distributions that manifest the unstable nature of the dynamics due to the strongly attractive SOC term. 

The structure of this work proceeds as follows. 
In Sec.~\ref{sec:2}, we describe the bosonic mixture and the eGPEs used to capture SOC QDs. Next, we discuss the SOC ground state QD configurations for different parametric variations together with the underlying excitation spectrum identifying the underlying stability properties in Sec.~\ref{sec:3}. Sec.~\ref{sec:3c} narrates the dynamical response of the droplets utilizing velocity perturbations and the analysis of quenches of the interaction or SOC parameters. We conclude and elaborate on future research directions in Sec.~\ref{sec:4}. Appendix~\ref{app:bdg} provides the matrix elements of the linearized eigenvalue problem. In Appendix~\ref{app_breath}, we describe the dependence of the droplet breathing frequency in terms of the SOC characteristics and in Appendix~\ref{stripe_quench}, we showcase the generation of stripe droplet branches after quenching the SOC wavenumber. 

\section{Beyond Mean-field model for SOC 
droplets}
\label{sec:2}

We consider a 1D pseudo-spin-$1/2$ bosonic gas experiencing strong transverse confinement such that the motion in these directions is frozen~\cite{hou2018momentum}. The involved spin states feature the same intracomponent repulsion $g_{\downarrow \downarrow}=g_{\uparrow \uparrow} \equiv g$ and intercomponent attraction of strength $g_{\uparrow\downarrow}$ which enables to enter the droplet regime in the presence of quantum fluctuations. 
The corresponding coupled set of eGPEs in dimensionless units~\cite{Tononi2019,Gangwar2022,Gangwar2023} reads 
\begin{subequations}
\label{eq:gpsoc:1}
\begin{align}
\mathrm{i} \partial_t \psi_{\uparrow}=\bigg[ &-\frac{1}{2}\partial_x^2-\mathrm{i} k_{L} \partial_x + \frac{\delta g}{2}\left(\lvert \psi_{\uparrow} \rvert^2 - \lvert \psi_{\downarrow} \rvert^2 \right) 
+ g \lvert \psi_{\uparrow} \rvert^2 \notag \\
&+ g_{\uparrow\downarrow} \lvert \psi_{\downarrow} \rvert^2 -\dfrac{g_{\text{LHY}}^{3/2}}{\pi}\sqrt{\lvert \psi_{\uparrow}\rvert^2 +\lvert\psi_{\downarrow}\rvert^2}\bigg] \psi_{\uparrow}+ \Omega \psi_{\downarrow}, \label{eq:gpsoc2-a} \\
\mathrm{i} \partial_t \psi_{\downarrow}=\bigg[ &-\frac{1}{2}\partial_x^2 +\mathrm{i} k_{L} \partial_x 
+ \frac{\delta g}{2}\left( \lvert \psi_{\downarrow} \rvert^2 - \lvert \psi_{\uparrow} \rvert^2 \right) 
+ g_{\downarrow\uparrow} \lvert \psi_{\uparrow} \rvert^2 \notag \\
&+ g \lvert \psi_{\downarrow} \rvert^2 -\dfrac{g_{\text{LHY}}^{3/2}}{\pi}\sqrt{\lvert \psi_{\uparrow}\rvert^2 +\lvert\psi_{\downarrow}\rvert^2}\bigg] \psi_{\downarrow}+ \Omega \psi_{\uparrow}. \label{eq:gpsoc2-b}
\end{align}
\end{subequations}
Here, $\psi_\uparrow$ ($ \psi_\downarrow$) denotes the 1D wave function of the spin-up (down) component, $k_L$ is the spin-orbit wavenumber, and $\Omega$ refers to the Rabi coupling frequency among the aforementioned spin states. {
Note that we consider zero detuning in our SOC model; consequently, only the stripe and zero-momentum phases can exist, while the plane-wave phase is excluded~\cite{li2012quantum, khamehchi2014measurement, chen2022elementary}.
} The assumption of equal intracomponent repulsions {and atom number per component leads to 
$\abs{\psi_{\uparrow}} = \abs{\psi_{\downarrow}}$}. For this reason, we choose to discuss below the features of the total wave function of the SOC system. Moreover, the strength of the LHY first-order quantum correction is $g_{\text{LHY}}=g$ while $\delta g = g_{\uparrow \downarrow}+ g$ is the mean-field balance point. 
Additionally, we employ the following condition on the system wave function~\cite{Astrakharchik2018} 
\begin{align}
\int\limits_{-\infty}^{\infty} 
\left[\lvert \psi_\uparrow \rvert^2 + \lvert \psi_\downarrow \rvert^2 \,\right] dx = 2N, \label{normalization}
\end{align}
where $2N$ is the normalization constant. 
It determines the normalized number of particles in the droplet (see also the discussion below). 

The energy of the system is measured in terms of $\hbar\omega_\perp$, with $\omega_\perp$ being the trap frequency in the transverse direction. 
In this sense, the characteristic length scale is set by the transverse harmonic oscillator length $a_{\perp}=\sqrt{\hbar/(m\omega_{\perp})}$ and time is expressed in units of $\omega_{\perp}^{-1}$. 
Also, $g$ = $2 \mathcal{N} a_{\uparrow \uparrow}/ a_{\perp}$ and $g_{\uparrow\downarrow} = 2 \mathcal{N} a_{\uparrow \downarrow} / a_{\perp}$, where $a_{\uparrow \uparrow}$ ($a_{\uparrow \downarrow}$) refers to the 3D intracomponent (intercomponent) s-wave scattering length which is tunable via Feshbach resonances~\cite{Semeghini2018}. 
$\mathcal{N}$ denontes the total particle number. 
The SOC wavenumber and intensity (Rabi coupling) are rescaled as $k_L \to \tilde{k}_L a_{\perp}$ and $\Omega \to \tilde{{\Omega}}/\omega_{\perp}$ respectively, while the wave function as $ \psi_{\uparrow, \downarrow} = \tilde{\psi}_{\uparrow, \downarrow} \sqrt{a_{\perp}}$. Here, the tilde quantities represent the dimensionfull ones.

Experimentally, our setup is realizable with a binary mixture consisting of the hyperfine states $\ket{\uparrow} \equiv \ket{F = 1, m_F = -1}$ and $\ket{\downarrow} \equiv \ket{F = 1, m_F = 0}$ of $^{39}$K ~\cite{Cabrera2018, Cheiney2018, Ferioli2019}. The validity of the 1D geometry is ensured by employing strong transverse confinement of frequency $\omega_\perp = 2\pi \times 800$Hz resulting in $a_\perp = 0.568\,\mu$m, while along the axial direction either a box potential or a relatively suppressed confinement e.g. $\omega_x = 2\pi \times 50$ Hz can be used~\cite{Cheiney2018}. Moreover, the total particle number in the droplets herein is assumed to take values in the interval $\mathcal{N}\sim [10^4,10^6]$, which is achieved by varying the normalization constant of Eq.~(\ref{normalization}) within the range $N=[1,100]$. On the other hand, by choosing the scattering length values $a_{\uparrow \uparrow} = a_{\downarrow \downarrow} = 0.5367 a_0/N$, where $a_0$ is the Bohr radius, a nearly negligible mean-field interaction (i.e. $\delta g \approx 0$) is attained leading to a dimensionless intracomponent coupling $g = g_{\mathrm{LHY}} = 1$. Turning to a finite $\delta g$, the intercomponent interaction takes values $g_{\uparrow\downarrow} = [-0.99, -0.6]$, allowing for $\delta g = [0.01, 0.4]$ that can be obtained with $a_{\uparrow \downarrow} = 1/N [-0.5313a_0, -0.322a_0]$.

The Rabi coupling frequency ($\Omega$) depends on the intensity of the underlying Raman lasers~\cite{lin2011spin,zhang2012collective}, and can be typically adjusted in the experiment within the interval $\tilde{\Omega} = 2\pi \times [0.8-32 ]$kHz. The latter corresponds to a dimensionless Rabi coupling frequency $\Omega = [1,40]$. Additionally, the wavenumber of the spin-orbit coupling ($k_L$) is tunable via either the laser's geometry or wavelength~\cite{lin2011spin, hou2018momentum, qu2013observation}. Below, we consider values of the dimensionless SOC wavenumber $k_L = [0.1, 3]$, related to variations of the laser wavelength from $\lambda_L = 25.23~\mu$m to $841$nm~\cite{lin2011spin}.

To address the ground state properties of the SOC droplet setting described by Eqs.~(\ref{eq:gpsoc2-a}) and (\ref{eq:gpsoc2-b}), we employ an imaginary time propagation method using a split-step Crank-Nicolson scheme~\cite{Muruganandam2009, Young2016, Ravisankar2021}. In the adopted dimensionless units, the considered box size corresponds to $L=307$ with spatial resolution $dx = 0.025$, and for the dynamics, a fixed time step $dt = 10^{-5}$ is utilized. A Gaussian initial ansatz with anti-symmetric profiles between the components, i.e., ${\psi}_{\uparrow}(x)=-{\psi}_{\downarrow}(-x)$ is used. 

\section{SOC droplet configurations and their spectrum}
\label{sec:3}
\subsection{Different droplet and stripe phases}
\label{sec:3a}

As explicated above, it is known that QDs of binary short-range bosonic mixtures, in the absence of SOC, assemble in different phases upon variations of the system parameters, namely the atom number~\cite{Astrakharchik2018, Parisi2020}, the mean-field interactions~\cite{Tylutki2020, Mistakidis2021} but also the trap strength~\cite{katsimiga2023solitary}. For instance, using fixed interactions and a larger particle number, the QD experiences a transition from a quasi-Gaussian soliton-like structure to an FT configuration~\cite{Astrakharchik2018}. On the other hand, it was recently demonstrated that the presence of SOC is responsible for spatial undulations, so-called stripes, of the droplet for different interactions and keeping the remaining system parameters fixed~\cite{Gangwar2023}. Below, we analyze emergent ground state properties of SOC droplets, especially focusing on the interplay between the involved mean-field interactions and the normalized atom number. 
\begin{figure}[!ht]
\centering\includegraphics[width=0.99\linewidth]{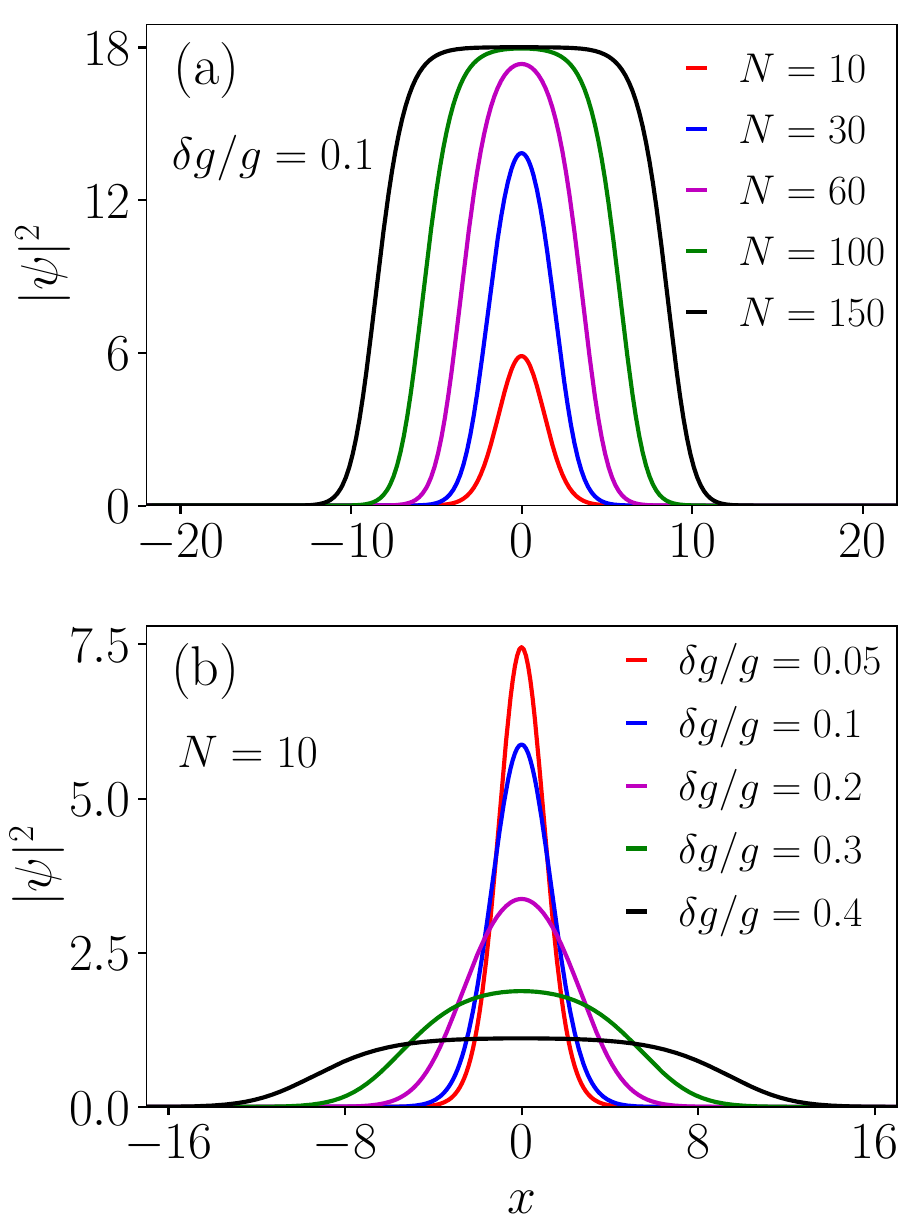}
\caption{Density deformation of a non-modulated QD with $\Omega=1$ and $k_L=0.5$. The total density is presented for (a) fixed mean-field interactions $\delta g/g=0.1$ and varying normalized atom number $N$ and (b) for constant $N=10$ and different $\delta g/g$. 
The droplet transforms from the Gaussian-like configuration to an FT one for larger $N$ featuring a saturation peak density in the FT regime. The transition threshold value dictated by $N$ decreases for larger $\delta g/g$. However, increasing $\delta g/g$ while $N$ held constant leads to a decrease of the peak density and a flattened profile, e.g. in the case of $N=10$ for  {$\delta g/g \gtrsim 0.3$}.}
\label{fig:denqdpw}
\end{figure}
\begin{figure*}[!ht] 
\centering\includegraphics[width=0.99\linewidth]{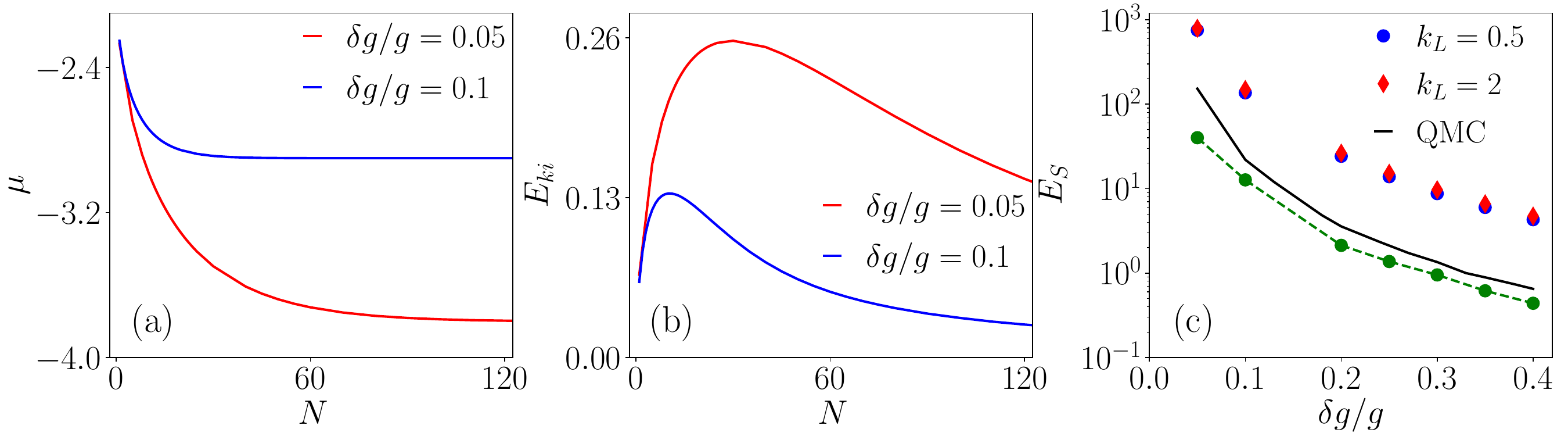}
\caption{Plots of (a) the total chemical potential, $\mu$, and (b) the total kinetic energy with respect to $N$ are shown for different interaction strengths $\delta g/g$ (see legend), with the other parameters being the same as in Fig.~\ref{fig:denqdpw}. The chemical potential shows a decreasing trend for larger $N$ in the quasi-Gaussian droplet regime and saturates above a threshold value $N_c$, beyond which an FT structure occurs. The aforementioned saturation value of $\mu$ becomes larger for increasing $\delta g/g$. The $E_{ki}$ attains its maximum at the transition $N_c$. (c) The behaviour of the surface energy $E_s$ in terms of $\delta g/g$ is shown. Upon fitting, it is found that the surface energy satisfies $E_s \sim (\delta g /g)^{-2.47}$.  {The green dotted line shows the difference of $E_s$ among the cases with SOC wave numbers $k_L=0.5$ and $k_L=2$.} The black solid line represents the many-body results reported for a binary short-range droplet using Quantum Monte Carlo Simulation (QMC) by Parisi and Giorgini \cite{Parisi2020}. {The observed deviations with the eGPE predictions are partly attributed to the presence of SOC (in the current setup) but also to residual beyond-LHY correlations that are taken into account within QMC.}}
\label{fig:mu}
\end{figure*}

Initially, we study the ground state density behaviour of QDs for varying normalized atom number but fixed interaction strengths, i.e. $\delta g/g=0.1$ with $g=1$, and $g_{\uparrow \downarrow} = -0.9$, as well as constant Rabi-coupling $\Omega=1$, and SOC wavenumber $k_L=0.5$, see Fig.~\ref{fig:denqdpw}(a). Notice that the choice of small $k_L=0.5$, satisfying $k_L^2 < \Omega$, ensures a non-spatially modulated (i.e. absence of stripe) structure in the droplet and thus the emergent configurations are similar to the ones in the absence of SOC~\cite{Ravisankar2021, Gangwar2022, Gangwar2023}. Indeed, the QD shows a transition from a quasi-Gaussian shape for small $N$, e.g.  {$N=10$}, towards an FT profile for larger $N$. Specifically, increasing $N$ leads to a gradual broadening of the width and the peak amplitude of the QD. After a certain normalized particle number threshold being here  {$N \gtrsim 100$}, the FT (peak density) of the QD saturates, and solely the width increases. The latter behaviour is a manifestation of the incompressible nature of the QD~\cite{Luo2020, Petrov2015, Petrov2016}, while the abovementioned threshold value depends, of course, on $\delta g/g$. 

Next, we examine the impact of the effective mean-field interaction parameter, namely $\delta g/g$, on the QD shape. 
The remaining system parameters are held fixed as in the previous case while $N=10$ is considered. Fig.~\ref{fig:denqdpw}(b) presents the respective QD density profiles for different $\delta g/g$, where since $g=1$ a larger $\delta g /g$ refers to a decreasing magnitude of intercomponent attraction $g_{\uparrow \downarrow}$. 
It can be readily deduced that for relatively small $\delta g/g$, for instance, $\delta g/g = 0.05$, the relevant density distribution is Gaussian-like. However, as $\delta g/g$ acquires larger values or equivalently, the intercomponent attraction becomes smaller in magnitude, and the QD size tends to increase, accompanied by simultaneous reduction of the peak density. The latter saturates for further increasing $\delta g/g$, see for example,  {$\delta g/g = 0.4$}, displaying an FT shape that persists for larger values of $\delta g/g$ where only the width of the QD keeps expanding. Overall, we once more observe that the droplet undergoes a transition from a quasi-Gaussian to an FT profile while attaining an intermediate broader Gaussian state, either for larger $N$ and fixed $\delta g/g$ or by increasing $\delta g/g$ for constant $N$.

The appearance of the three above-described structural regimes for distinct $N$ indicates the decisive competition between the individual energy contributions. In general, the existence of a self-bound state can be attributed to the balance between destabilizing and stabilizing terms in the system. For example, the attractive SOC and LHY terms attempt to destabilize the droplet, while the repulsive mean-field and kinetic energies could assist its stability. Indeed, at low $N$, the kinetic energy dominates over the mean-field one, resulting in a narrow Gaussian-shaped QD. Meanwhile, at intermediate $N$, the mean-field energy surpasses the kinetic one, leading to broader Gaussian-like droplets. Following a further increase of $N$, there is a corresponding increase in the repulsive mean-field energy and vanishing kinetic energy, ultimately leading to an FT QD structure~\cite{Astrakharchik2018}.

We confirm these intriguing characteristics of the distinct droplet states by examining the competition of the system's energy terms. As a first step, we provide in Fig. \ref{fig:mu} the total chemical potential and the underlying kinetic energy of both components as a function of $N$. It is evident that $\mu$ becomes more negative for larger $N$, and eventually, depending on the value of $\delta g/g$, it tends to a constant value [Fig. \ref{fig:mu}(a)]. For instance, in the case of $\delta g/g=0.05$, the saturation of $\mu$ takes place at a higher $N$ compared to $\delta g/g=0.1$. This implies that for larger $N$, the system attains a stronger bound state. Conversely, the kinetic energy initially increases as long as the QD refers to a narrow Gaussian; it reaches a maximum when the droplet acquires a broader Gaussian distribution and decreases in the FT region [Fig. \ref{fig:mu}(b)].

To further understand the modifications of the QD distribution, it is quite relevant to explore the so-called surface energy term. Recall that at large $\delta g/g$, the droplet reaches an FT profile with its density maximum being insensitive to larger $\delta g/g$, and only its width becomes wider. The surface energy of the QD, in 1D, can be estimated through the standard liquid drop model, namely expressing the total energy per particle within the large $N$ limit \cite{Astrakharchik2018, Parisi2020} as follows
\begin{align}
 \dfrac{E}{N}=E_v+\dfrac{E_s}{N}. 
\end{align} 
The first term, $E_v$, represents the bulk energy, and the second one is the correction due to the finite atom number, which scales as $1/N$ and depends on the surface energy coefficient $E_s$. Determining the energy per particle, $E/N$, for fixed $\delta g/g$ yields that it obeys a linear trend with increasing $1/N$. Since the slope of $E/N$ is proportional to $E_s$, we next determine the latter with respect to $\delta g/g$  {for the SOC wave number $k_L=0.5$ as well as $k_L=2$ (stripe phase) of the droplet}, see Fig.~\ref{fig:mu}(c). It is found that $E_s$ features a power-law dependence on $\delta g/g$ with an exponent of  {$-2.47$}. This behaviour signals the reduction of $E_s$ in the FT region and is attributed to the respective incompressible character.  {The difference of $E_s$ between $k_L=0.5$ and $k_L=2$ is significant at relatively small $\delta g /g$, and it decreases for $\delta g /g>0.25$, where the FT background becomes more prominent, see in particular the dotted green line in Fig.~\ref{fig:mu}(c).} 


\begin{figure}[!ht] 
\centering\includegraphics[width=0.99\linewidth]{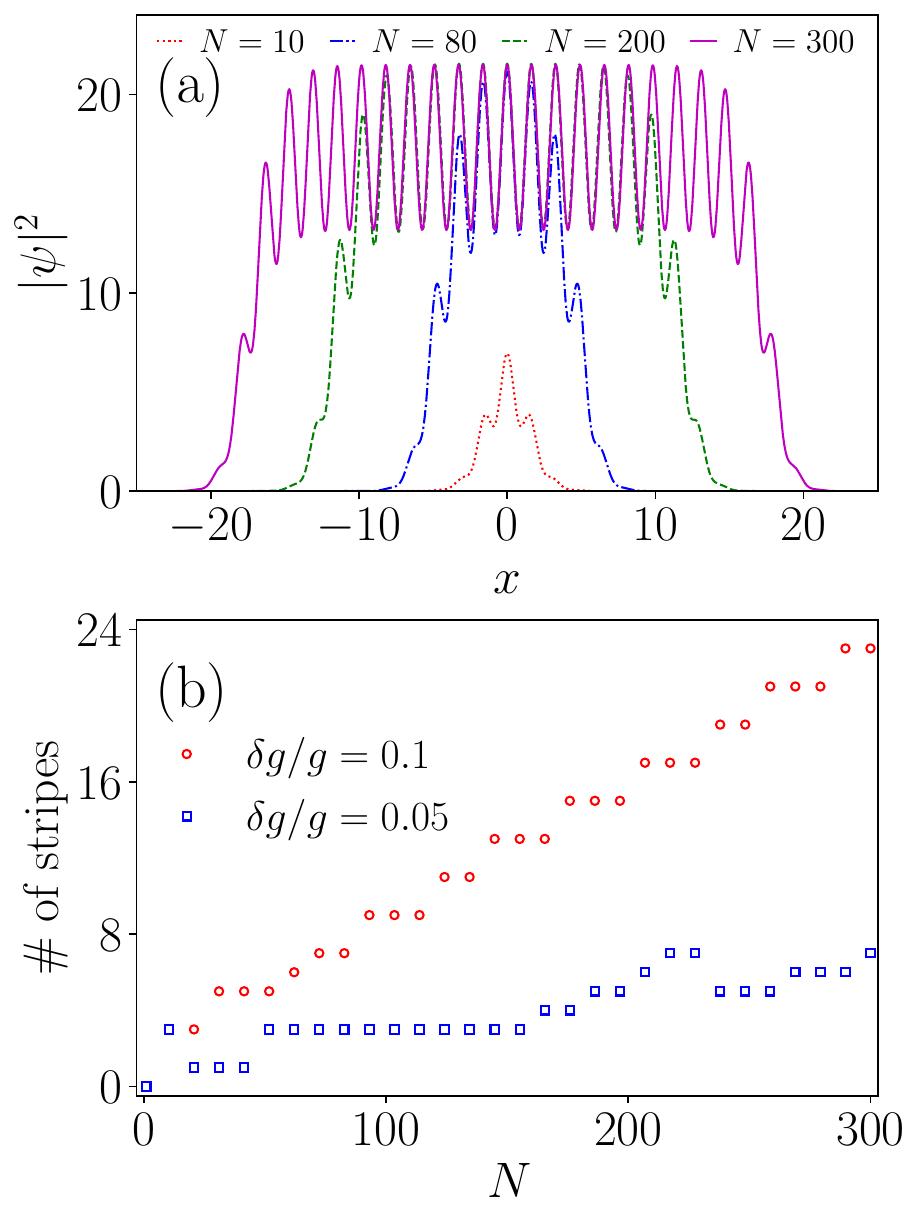} 
\caption{(a) Total density profiles of the quantum stripe droplet at $\delta g/g=0.1$ for several values of $N$ (see legend). The droplet configuration accommodates a larger number of stripes as $N$ increases, and eventually, its background flattens. (b) The number of resulting stripes as a function of the normalized atom number $N$ for various interactions $\delta g/g$ (see legend). Other system parameters are $g=1$, $\Omega=1$, and $k_L=2$.}
\label{fig:denqdsw}
\end{figure}

Subsequently, we employ a larger SOC wavenumber, such as $k_L=2$, which enforces significant spatial undulations in the droplet density and thus allows us to access the stripe phase~\cite{Gangwar2022}. For simplicity, we use $\Omega = 1$, and we examine the ground state of the stripe QD as a function of $N$ for a fixed $\delta g /g=0.05$, as depicted in Fig.~\ref{fig:denqdsw}(a). Apparently, the utilized SOC wavevector imprints stripe patterns in the droplet density almost irrespective of $N$. However, the number of these stripes crucially depends on $N$ and, in particular, increases for larger $N$, as shown in Fig.~\ref{fig:denqdsw}(b) as well. This is expected, as an increasing $N$ leads to a wider background, which can naturally host more stripes. Interestingly, beyond an interaction-dependent normalized particle number threshold (e.g., for $\delta g /g=0.05$, it refers to  {$N \gtrsim 200$}), the peak density of the central stripes attains that of the bulk density and remains unaltered for larger $N$. Similar to the non-modulated droplet phase, only the width (size) of the bulk density increases for higher $N$, while the central distribution features a flattened behaviour. Therefore, the number of stripes in the droplet continues to increase. This is explicitly shown in Fig.~\ref{fig:denqdsw}(b) for distinct interactions. Accordingly, the number of stripes is also enhanced at a given $N$ (especially for $N>10$) for larger $\delta g/g$, which favours a more spatially extended background.

\subsection{Excitation spectrum of the SOC droplet phases}
\label{sec:3b}

To extract the excitation spectrum and analyze the stability properties of the previously discussed SOC droplet solutions, we perturb the latter according to 
\begin{align}\label{eq:gpsoc:2}
\Psi_j(x , t)= \mathrm{e}^{-\mathrm{i} \mu_j t} \left[\psi _j (x) + u_j (x) e^{-i \omega t }+v^*_j (x) \mathrm{e}^{\mathrm{i} \omega^* t }\right]. 
\end{align}
 {In the above expression, $\psi_j$  and $\mu_j$ denote the chemical potentials and the complex ground state wave functions of the $j = (\uparrow, \downarrow)$ components. Furthermore, $u_j$ and $v_j$ denote the Bogoliubov amplitudes (i.e., the resulting eigenvectors), while $\omega$ refers to the eigenfrequency of the perturbation. Upon substituting the ansatz of Eq.(\ref{eq:gpsoc:2}) into the eGPE [Eq.(\ref{eq:gpsoc:1})], we obtain the following linearized eigenvalue problem:}
\begin{align}
\mathcal{L}
\begin{pmatrix}
u_\uparrow \\ v_\uparrow \\ u_\downarrow \\ v_\downarrow
\end{pmatrix}
= \omega 
\begin{pmatrix}
u_\uparrow \\ v_\uparrow \\ u_\downarrow \\ v_\downarrow
\end{pmatrix},\label{eq:bdgEigenval}
\end{align}
where the matrix 
\begin{align*}
\mathcal{L}=
\begin{pmatrix}
 f_1 & Y_1 &Z_1 & Z_2 \\
-Y_1^* & -f_1^* &-Z_2^* & -Z_1^* \\
 Z_1^* & Z_2 & f_2 &Y_2 \\
-Z_2^* & -Z_1 & -Y_2^* & -f_2^*  
\end{pmatrix}.
\end{align*}
We provide the explicit values of the matrix elements in Appendix~\ref{app:bdg}. Using the normalization condition $\int\left(\lvert u_j\rvert ^2-\lvert v_j^* \rvert^2 \right)dx=1$, we obtain the eigenspectrum by numerically solving $\mathrm{det}(\,\mathcal{L})=0$. 
\begin{figure}[!ht]
\centering\includegraphics[width=0.99\linewidth]{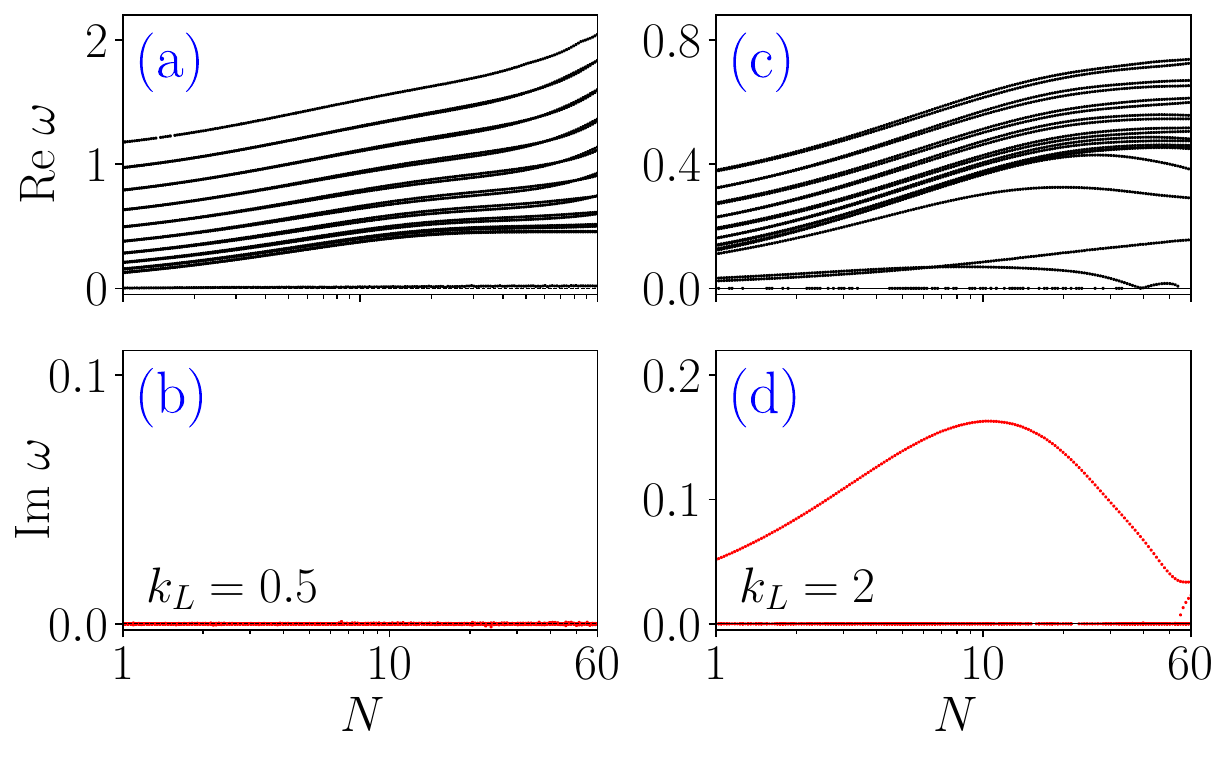}
\caption{ {The real (upper panels), Re $\omega$, and the imaginary (lower panels), Im $\omega$, parts of the first twenty lower lying eigenvalues of Eq.~(\ref{eq:bdgEigenval}) as a function of $N$ for (a), (b) $k_L=0.5$ and (c), (d) and $k_L=2$ (stripe phase). 
Here, the parameters $\delta g/g=0.1$ and $\Omega=1$ are kept fixed. Apparently, when $k_L=0.5$ the eigenvalues remain purely real, indicating the dynamical stability of the SOC droplet. 
However, imaginary contributions of the eigenvalue exist for $k_L=2$ 
independently of $N$, 
suggesting the dynamical instability of the stripe droplet.}}
\label{fig:bdg-spectrum1}
\end{figure}
\begin{figure}[!ht]
\centering\includegraphics[width=0.99\linewidth]{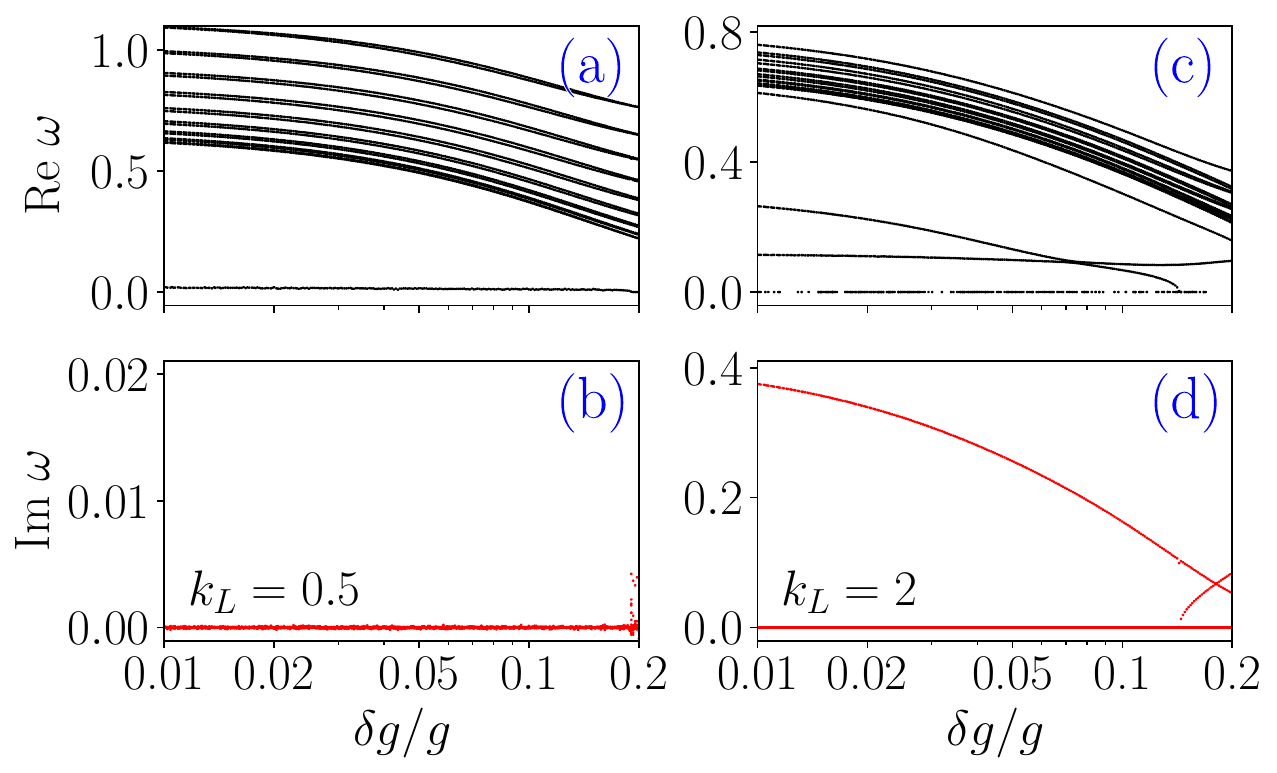}
\caption{ {(Upper panels) Real, Re $\omega$, and (lower panels) imaginary, Im $\omega$, parts of the eigenvalues of Eq.~(\ref{eq:bdgEigenval}) with varying $\delta g/g$ for (a), (b) $k_L=0.5$, and (c), (d) $k_L=2$. In all cases $N=10$ and $\Omega=1$. For $k_L=0.5$ the spectrum remains purely real, implying the dynamical stability of the droplet for all $\delta g/g$, while for $k_L=2$ the presence of imaginary eigenvalues evinces the dynamical instability of the stripe droplet. } }
\label{fig:bdg-spectrum2}
\end{figure}
 {
Figure~\ref{fig:bdg-spectrum1} shows the variation of the real and the imaginary parts of the low lying eigenvalue ($\omega$) with respect to $N$ for different SOC wave numbers, namely $k_L=0.5$ [Fig.~\ref{fig:bdg-spectrum1}(a)] and $k_L=2$ [Fig.~\ref{fig:bdg-spectrum1}(b)], with fixed $\Omega=1$.
Focusing on $k_L=0.5$, we observe that the eigenvalues remain purely real (with the imaginary part being zero) regardless $N$, i.e. for both the Gaussian and the FT droplet phases. 
This behavior affirms the dynamically stable nature of the droplet as it has been shown already for the 
binary short-range interacting droplet setting without SOC~\cite{Tylutki2020, katsimiga2023interactions, katsimiga2023solitary}. 
However, for the stripe phase ($k_L=2$) of the droplet, the eigenvalues exhibit a non-negligible 
imaginary contribution for all $N$ [cf. Fig.~\ref{fig:bdg-spectrum1} (c) and (d)], indicating the involvement of dynamical instabilities in this case. 
In a similar vein, analyzing 
the eigenvalues for $k_L=0.5$ and $k_L=2$ by keeping the normalization of the droplet fixed ($N=10$) and varying $\delta g/g$, we find that the eigenvalues remain purely real in the former case and are complex in the latter phase, as depicted in Fig.~\ref{fig:bdg-spectrum2}. 
Interestingly, an increase in $\delta g/g$ results in the decrease of the magnitude of the real and imaginary parts of the eigenvalue for $k_L=0.5$ and $k_L=2$, respectively. 
This feature of $\omega$ for increasing $\delta g/g$ can be attributed to the accompanied  decrease of the chemical potential 
also realized for the binary droplet in the absence of SOC~\cite{katsimiga2023interactions}. 
A more detailed analysis of the underlying instabilities and their effects as well as their dependence on the remaining system parameters is a fruitful prospect for future work. Here, also the case of nonzero quasi-momenta can be  taken into account with multiplying the term $~\sim e^{ikx}$ in the linearization ansatz of Eq.~(\ref{eq:gpsoc:2}) which will reveal the presence of the zero-momentum and plane wave ~\cite{li2012quantum, khamehchi2014measurement, chen2022elementary} besides the stripe one that we find in the present analysis. 
} 
\color{black}%
\section{Dynamics of SOC quantum droplets}
\label{sec:3c}

\begin{figure*}[!htp]
\centering\includegraphics[width=0.99\linewidth]{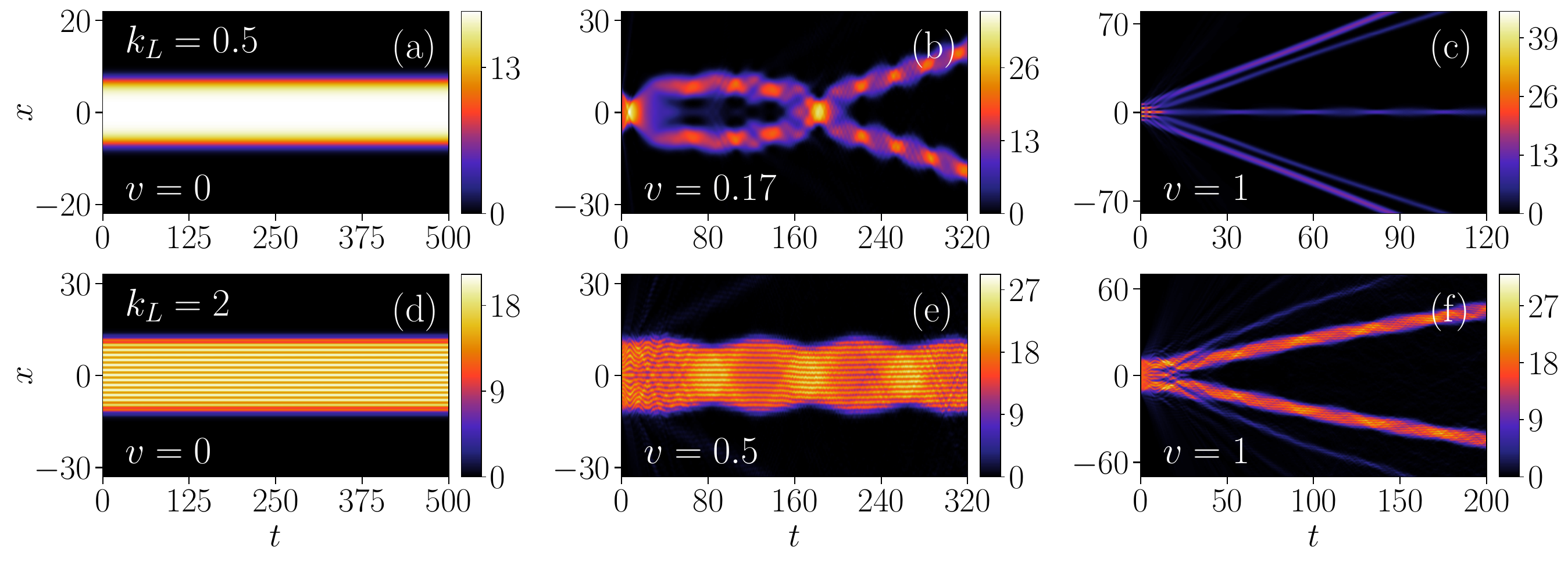} 
\caption{Total density evolution of different SOC QDs subjected at $t=0$ to velocity perturbations (a), (d) $v=0$, (b)  {$v=0.17$}, (c), (f)  {$v=1$} and (e) $v=0.5$. The droplets are characterized by SOC wavenumber (a)-(c) $k_L=0.5$ for  {$N=120$} and (d)-(f) $k_L=2$ for  {$N=200$}. They are initiated in the ground state where $g=1$, $g_{\uparrow\downarrow}=-0.9$ ($\delta g/g = 0.1$) and $\Omega =1$. When $v=0$, the QD propagates undistorted, while for finite velocity, it features a gradual transition from a breather [panel (e)] to moving [panels (c), (f)], i.e. multiple fragmented ones due to the dominant kinetic term contribution. The number of fragmented QDs increases for larger $N$.}
\label{fig:dynapw}
\end{figure*}

Having analyzed the ground state properties of the droplet phases under the influence of SOC, we next proceed to study the dynamical response of these entities subjected to different external perturbations. These include (i) imprinting an initial velocity to the droplet through a uniform change in the phase of its ground state wave function, (ii) following a quench of the mean-field interaction parameter, or (iii) applying a quench to the Rabi-coupling. The time evolution of both the standard droplet and the striped one is examined. As in the recently studied case~\cite{Gangwar2022} of vanishing mean-field interactions ($\delta g=0$), dominant emergent features are breathing motion and dynamical fragmentation of the droplet into multiple fragments as well as droplet splitting for increasing magnitude of its initial velocity. 

\subsection{Impact of the initial velocity: transition from breather to fragmented droplets}

Here, we explore the response of droplets subjected to a phase shift on the ground state wave function of the individual components, namely $\psi(x,t=0)=\psi_{\uparrow} \exp{(-ivx)} + \psi_{\downarrow} \exp{(+ivx)}$. As a case example, a FT droplet containing $N=60$ atoms and characterized by $\Omega=1$, $k_L=0.5$, $g=1$, and $g_{\uparrow\downarrow}=-0.9$ ($\delta g/g=0.1$) is employed. The resultant dynamics of such a droplet are presented in Fig.~\ref{fig:dynapw}(a)-(c) for different velocities, namely $v=0$,  {$v =0.17$, and $1$}. In the case of $v=0$, the droplet remains unperturbed throughout the evolution, thus confirming its stable nature [Fig.~\ref{fig:dynapw}(a)]. However, for finite velocities, the droplet response is drastically modified [see Figs.~\ref{fig:dynapw}(b) and (c)]. For instance, at  {$v=0.17$}, the phase perturbation triggers internal excitations of the droplet that are predominantly of breathing type, as can be seen in Fig.~\ref{fig:dynapw}(b). Interestingly, the breathing droplet splits into two parts around  {$t\sim 40$}, which then recombine at  {$t\sim 180$} into a single droplet. In the vicinity of the droplet splitting, the kinetic energy contribution dominates. Afterwards, $t\sim 180$, the single excited droplet separates again into two oppositely moving and not equally mass ones. Notice here that a smaller normalized atom number which is related to a quasi-Gaussian droplet undergoes solely a breathing motion. A further increase of the initial velocity, e.g.  {$v=1$} depicted in Fig.~\ref{fig:dynapw}(c), leads to a dynamical fragmentation process of the FT droplet. Indeed, the fragmented density parts here consist of four outer counterpropagating droplets featuring breathing-like oscillations and an inner quasi-stationary droplet residing around $x=0$. The number of fragmented droplets depends strongly on the involved normalized particle number, e.g. it is three for  {$N=80$} and five for  {$N=120$}.

Turning to the effect of velocity perturbations on the stripe droplet state we observe a somewhat modified response [see Fig.~\ref{fig:dynapw}(d)-(f)]. Here, the employed ground state stripe droplet waveform is characterized by  {$N=200$}, $\Omega=1$ and $k_L=2$, while the interaction parameters are the same as in the above discussion. For these parameters, the respective ground state configuration turns out to contain nineteen stripes. As expected, for $v=0$, the stripe droplet (independently of $N$) is unchanged in both size and shape in the course of the evolution verifying the stability of the state [Fig.~\ref{fig:dynapw}(d)]. However, for finite velocities (e.g. $v=0.5$), the droplet stripe experiences internal excitations corresponding to the vibrational motion of the participating stripes [see Fig.~\ref{fig:dynapw}(e)]. This, in turn, results in an overall breathing motion of the entire background accompanied by the emission of density portions that become more pronounced for a higher number of atoms. This dynamical response will be referred to in the following as \textit{breather-like} droplet; it occurs also for the standard droplet described above (not shown). Importantly, upon considering higher velocities, namely  {$v=1$} here, there is a transition to the so-called \textit{fragmented droplet} dynamical stage as can be clearly seen in Fig.~\ref{fig:dynapw}(f). This fragmentation process is associated with the dominance of the attractive SOC energy term, and it means that the initial droplet breaks into several smaller ones whose specific number and behaviour are dictated by the particle number. For instance, when  {$N=200$} at short evolution times ($t<20$), the strong perturbation causes stripe collisions and coalescence. As a result, there is a burst of density emission (radiation) from the ``surface" of the droplet accompanied by the nucleation of four counterpropagating stripe droplets featuring breathing excitation. These outer moving droplet stripes retain their nature for long evolution times. In contrast, for  {$N=180$} (not shown), the perturbation-induced initial stripe collisions lead to the droplet breaking into three parts, with the central one resembling a breather stripe droplet and the outer ones travelling outwards. 
\begin{figure}[!htp]
\centering\includegraphics[width=0.99\linewidth]{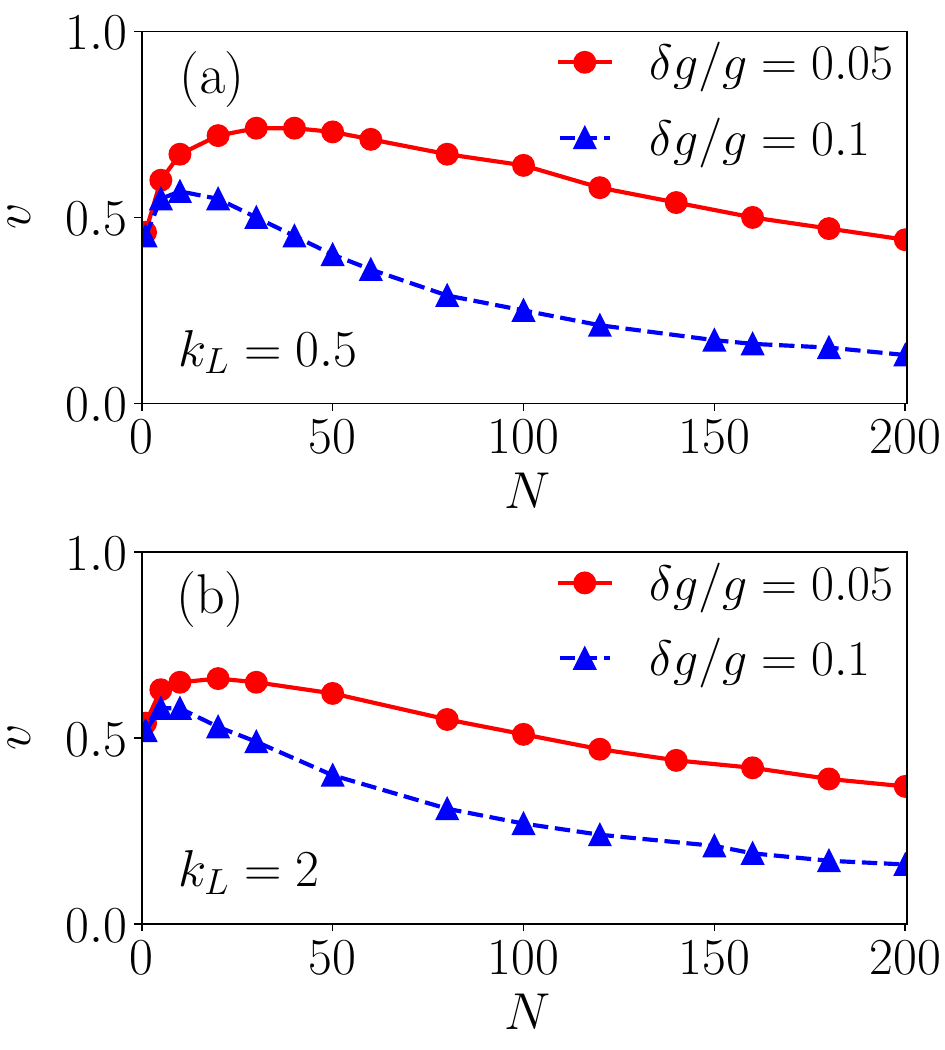} 
\caption{Critical velocity, $v_c$, of the droplet above which it dynamically transforms from a breather to a fragmented (moving) one with respect to $N$ and different $\delta g/g$ (see legends). The droplet is characterized by $\Omega=1$ and a SOC wavenumber (a) $k_L=0.5$ (non-modulated solution) and (b) $k_L=2$ (stripe state). 
The $v_c$ decreases with increasing $N$ reaching an interaction-dependent maximum where the droplet is more stable against velocity perturbations. Here $v_c$ corresponds to the minimum of the total energy.}
\label{fig:cricvel}
\end{figure}

\begin{figure*}[!htp]
\centering\includegraphics[width=0.99\linewidth]{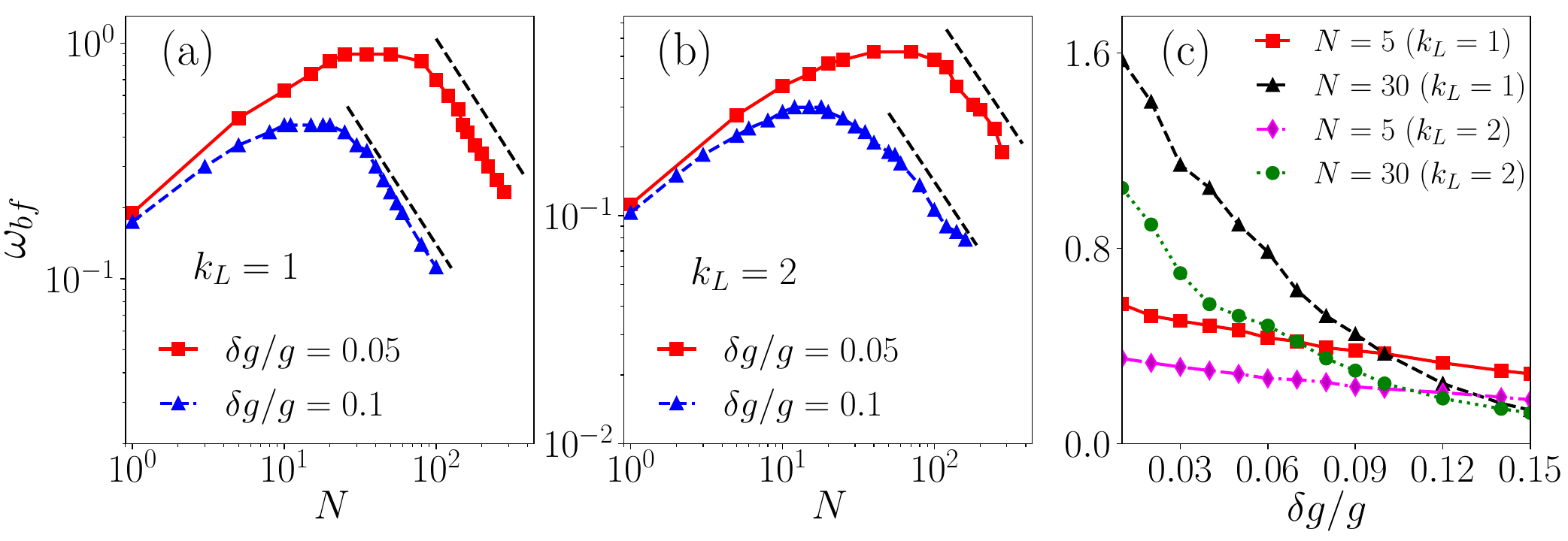} 
\caption{(a) Behavior of the droplet breathing frequency ($\omega_{bf}$) with respect to $N$ for (a) $k_L=1$ and (b) $k_L=2$ as well as different interactions $\delta g/g$ (see legends). The ground state droplet characterized by $g=1$ and $\Omega=1$ is perturbed with the initial velocity $v=0.05$. The dotted lines provide a guide to the eye illustrating the behaviour of $\omega_{bf}\sim N^{-1}$ at large $N$ where FT droplets form. Also, $\omega_{bf}$ is generically smaller for stripe droplets as can be inferred by comparing panels (a) and (b). (c) $\omega_{bf}$ as a function of $\delta g/g$ for distinct $N$ and $k_L$ (see legend).}
\label{fig:brefre1}
\end{figure*}

The transition from the breather to the fragmented or moving droplet regime is naturally characterized by a critical velocity, $v_c$, above which this structural deformation occurs. This critical velocity is determined by identifying the velocity at which the total energy at $t>0$ attains its minimum value for a given set of parameters. Recall that the total energy of a stripe ground state droplet decreases for a larger amount of stripes. The behaviour of $v_c$ for $k_L=0.5$ (droplet) and $k_L=2$ (stripe droplet) as a function of the normalized atom number $N$ for $\delta g/g=0.05,0.1$ is depicted in Fig.~\ref{fig:cricvel}(a) and (b), respectively. In both cases, i.e. striped or not droplet, $v_c$ has an increasing trend with $N$, showing a maximum at the normalized particle number at which the transition from the broader Gaussian to the FT droplet state takes place. Eventually, $v_c$ decreases as the FT regime is reached. For example, in the case of $k_L=0.5$ for $\delta g/g=0.05$ ($\delta g/g=0.1$), the critical velocity exhibits a maximum around  {$N \sim 30$ ($N \sim 10$)} and afterwards decreases. This particular feature indicates that the droplet at the maximum value of $v_c$ is more stable against velocity perturbations as compared to droplets with different sizes~\cite {Astrakharchik2018}.


\subsection{Droplet breathing frequency} 

Next, we explore the parametric dependence of the droplet breathing frequency on the involved atom number and interactions. The breathing frequency, $\omega_{bf}$, is identified in the spectrum of the time-evolved width of the droplet wavepacket~\cite{mistakidis2021formation}, i.e. $\braket{\psi\vert x^2(t)\vert \psi}$, after a weak velocity perturbation with $v=0.05$. The cases of both the quasi-Gaussian with $k_L=1$ [Fig.~\ref{fig:brefre1} (a)] and the stripe droplet with $k_L=2$ [Fig.~\ref{fig:brefre1} (a)] are studied for completeness. It is apparent that $\omega_{bf}$ exhibits similar characteristics in both regimes. In particular, independently of $\delta g/g$, the breathing frequency features an overall increasing trend with $N$ until it reaches a maximum and subsequently decreases at larger $N$. The value of $N$, at which $\omega_{bf}$ maximizes, appears to be smaller for stronger $\delta g/g$. Also, $\omega_{bf}$ is higher for lower $\delta g/g$ and fixed $N$ meaning that FT structures possess a smaller $\omega_{bf}$ in accordance with observations for non SOC droplet configurations~\cite{Parisi2020, Englezos2023}. Furthermore, the decreasing behaviour of $\omega_{bf}$ at large $N$ is to a good approximation inversely proportional to the atom number, namely it scales as $N^{-1}$, see also the respective fitted dotted black lines. This is consistent with the fact that the droplet displays phonon-like excitations once it reaches an FT background, as it was also reported in \cite{Astrakharchik2018}. 

Another interesting observation is that the magnitude of $\omega_{bf}$ for a specific $N$ and $\delta g/g$ is lower when increasing $k_L$, as can be deduced by comparing Fig.~\ref{fig:brefre1}(a) and (b). To further elucidate this issue, we select $N=5$ and $N=30$ for both $k_L=1$ and $k_L=2$ and investigate $\omega_{bf}$ in terms of $\delta g/g$, as shown in Fig.~\ref{fig:brefre1}(c). Evidently, for $N=5$, the breathing frequency decreases almost linearly with $\delta g/g$, irrespective of $k_L$. However, for large $N$, $\omega_{bf}$ shows a non-monotonic behavior. It is higher at smaller $\delta g/g$ and reduces at a faster rate with $\delta g/g$ compared to the $N=5$ case. Also, at large values of $\delta g/g$, $\omega_{bf}$ is larger for $N=5$ than for $N=30$, independently of $k_L$, since, as explained above, Gaussian structures with smaller widths have an enhanced $\omega_{bf}$.

\subsection{Intercomponent interaction quench dynamics of the SOC quantum droplet}
\label{sec:3d}

Our next focus is to explore the non-equilibrium dynamics of the droplet after a sudden change of the interaction parameter $\delta g/g$. To be more concrete, we keep the intracomponent repulsion constant at $g=1$, and we quench the intercomponent interaction $g_{\uparrow\downarrow}$ to either larger or smaller attractive strengths compared to the original value of $g_{\uparrow\downarrow} = -0.9$, corresponding to $\delta g/g = 0.1$. The response of the system is exemplary and monitored through the total density of the SOC setting with $k_L=2$ and $\Omega=1$. Similar patterns to the ones presented below also occur for $k_L=0.5$.

\begin{figure*}[!htp]
\centering\includegraphics[width=0.99\linewidth]{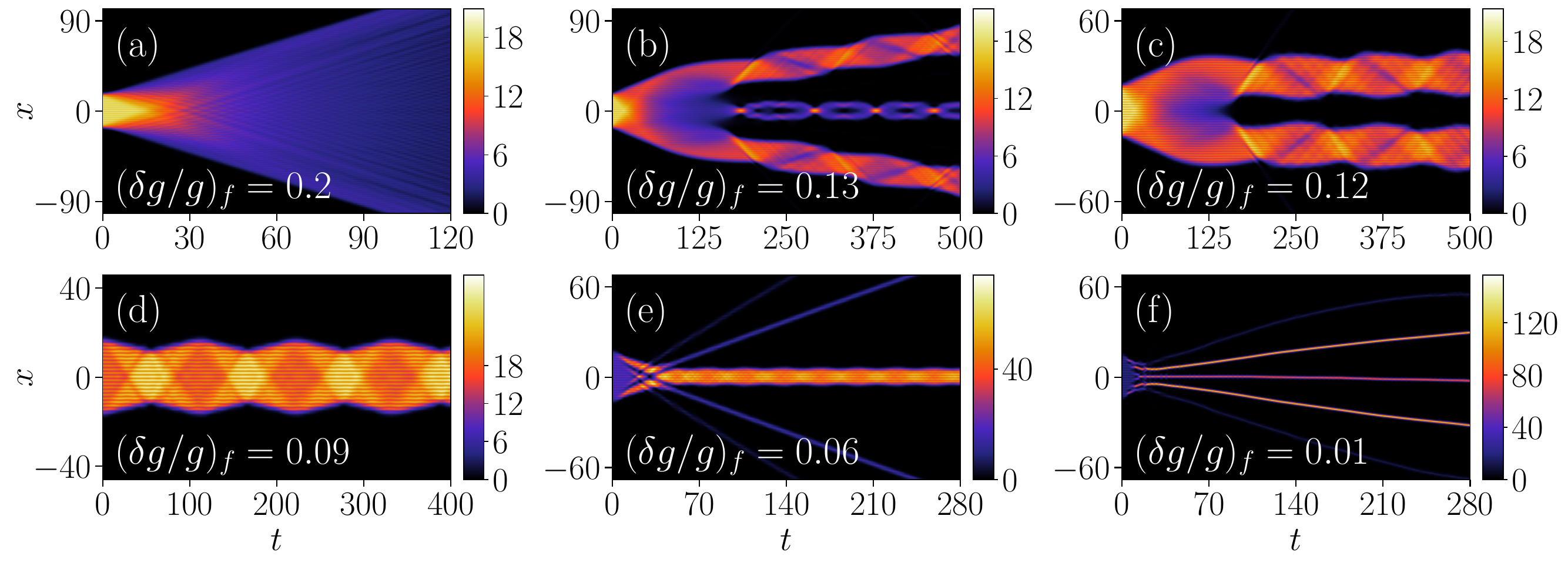}
\caption{Temporal evolution of the stripe droplet density following an intercomponent interaction quench from $(\delta g / g)_{{\rm in}}=0.1$ to different values of $(\delta g /g)_{{\rm f}}$ (see legends). 
The initial ground state is prepared with $\Omega =1$, $k_L = 2$,  {$N=280$}, $g=1$, and $g_{\uparrow\downarrow}=-0.9$ ($\delta g/g = 0.1$). Quenching to less attractive intercomponent interactions leads to droplet expansion [panel (a)] and splitting [panels (b), (c)]. In contrast, for quenches to stronger attractions, the droplet either performs a breathing motion [panel (d)] or fragments [panels (e), (f)]. The explicit post-quench interaction values correspond to (a) $g_{\uparrow \downarrow}=-0.8$ ($\delta g/g = 0.2$), (b) $g_{\uparrow \downarrow}= -0.87$ ($\delta g/g = 0.13$), (c) $g_{\uparrow \downarrow} = -0.88$ ($\delta g/g = 0.12$), (d) $g_{\uparrow \downarrow} = -0.91$ ($\delta g/g = 0.09$), (e) $g_{\uparrow \downarrow} = -0.94$ ($\delta g/g = 0.06$) and (f) $g_{\uparrow \downarrow}= -0.99$ ($\delta g/g = 0.01$). Similar features can also be observed for $k_L=0.5$.}
\label{fig:quenbetasw}
\end{figure*}

Initially, we examine the effect of quenches towards larger values of $\delta g/g$, which essentially correspond to less attractive intercomponent couplings, as shown in Fig.~\ref{fig:quenbetasw}(a)-(c). In the ground state of the system, this means that the solution tends to the FT regime, as illustrated, for instance, in Fig.~\ref{fig:denqdpw}(b). Naturally, this type of quench, due to the reduced post-quench attraction, leads, generally, to an overall expansion of the stripe droplet, but the explicit patterns depend on the specific interaction value. This is evident for all three $(\delta g/g)_{{\rm f}}$ values presented in Fig.~\ref{fig:quenbetasw}(a)-(c). In particular, a relatively large post-quench amplitude, e.g., $(\delta g/g)_{{\rm f}}=0.2$, results in a continuum expansion of the droplet in terms of its width and a reduction of its amplitude since the FT is dynamically attained [Fig.~\ref{fig:quenbetasw}(a)]. Here, the kinetic and attractive SOC energy terms are the most prominent ones and thus responsible for the observed dynamics. Notice that such a droplet expansion can be equally observed in short-range attractive interacting bosonic mixtures, as reported in Ref.~\cite{mistakidis2021formation}.

However, a slightly smaller quench amplitude, e.g. $(\delta g/g)_{\rm f} = 0.13$ or $0.12$, is accompanied by richer dynamical patterns (see Fig.~\ref{fig:quenbetasw}(b), (c)). Besides the expansion of the background, a droplet-splitting process takes place, which is less pronounced for interactions closer to the original ones. The splitting events are caused by the progressively dominant contribution of the attractive LHY energy term over the remaining ones in time. For instance, at $(\delta g/g)_{\rm f} = 0.13$, we observe the generation of three droplets from the initial structure. The two outer ones are counterpropagating, having equal velocities, and they simultaneously experience breathing motion, while the inner droplet remains at the center, showing splitting and merging events [Fig.~\ref{fig:quenbetasw}(b)]. We remark that, in the process, excitations also propagate from the surface to the bulk. Accordingly, for a comparatively smaller increment of the interactions, e.g., $(\delta g/g)_{\rm f} = 0.12$, solely two counterpropagating droplets with less velocity are emitted [Fig.~\ref{fig:quenbetasw}(c)]. These emitted droplets, being closer to each other, especially as compared to the one in the $(\delta g/g)_{\rm f} = 0.13$ quench, feature an attraction soon after their creation. This behaviour is in line with one of the closely placed short-range droplets whose effective attractive force can be explained by constructing an effective particle picture\cite{katsimiga2023interactions}.

On the other hand, applying interaction quenches towards lower $(\delta g/g)_{{\rm f}}$ values or equivalently more attractive intercomponent interactions than the pre-quench one favours, in general, the dynamical fragmentation of the stripe droplet, see Fig.~\ref{fig:quenbetasw}(d)-(f). 
Referring to the respective ground state configuration, the waveform of the post-quench interaction is a Gaussian with shrunk width. Specifically, for small quench amplitudes, the stripe droplet experiences a breathing motion with the individual stripes featuring vibrational modes [Fig.~\ref{fig:quenbetasw}(d)]. This response changes drastically for larger post-quench attractions where at short times, the destructive interference of the involved stripes generates radiation and structures resembling dark solitons at the vicinity of $x=0$. These structures suffer further collisions leading to eventual fragmentation into multiple highly localized droplets, a process that is more pronounced for smaller $(\delta g/g)_{{\rm f}}$, see Fig.~\ref{fig:quenbetasw}(e) and (f). Such violent fragmentation events are inherently related to the prevailing strong attractive LHY energy contribution. We remark that at these suppressed $\delta g$ values, where mean-field interactions cancel out, and solely quantum fluctuations are present, the system is close to the LHY fluid~\cite{Joergensen2018}, which has been experimentally realized~\cite{Skov2021} in the absence of SOC and occurs at $\delta g=0$. A similar droplet breaking mechanism into multiple ones was also found to occur in the context of dipolar droplets \cite{Edmonds2020}.

\subsection{Effect of the Rabi-coupling}
\label{sec:3e}

A natural next question that arises concerns the impact of the SOC parameters, i.e. the Rabi-coupling and the wavenumber, on the dynamical stability of the QD. Here, we first consider sudden changes of the Rabi-coupling and, in particular, divide our analysis into two parts: i) examine the time-evolution upon quenching from $\Omega=0$ to a finite one $\Omega_{{\rm f}}>0$ and ii) quench $\Omega$ from a finite value to a larger one. We remark that keeping all other system parameters constant and increasing $\Omega$ is associated with a decrease of the ground state energy towards more negative values. 

\begin{figure*}[!htp]
\centering\includegraphics[width=0.99\linewidth]{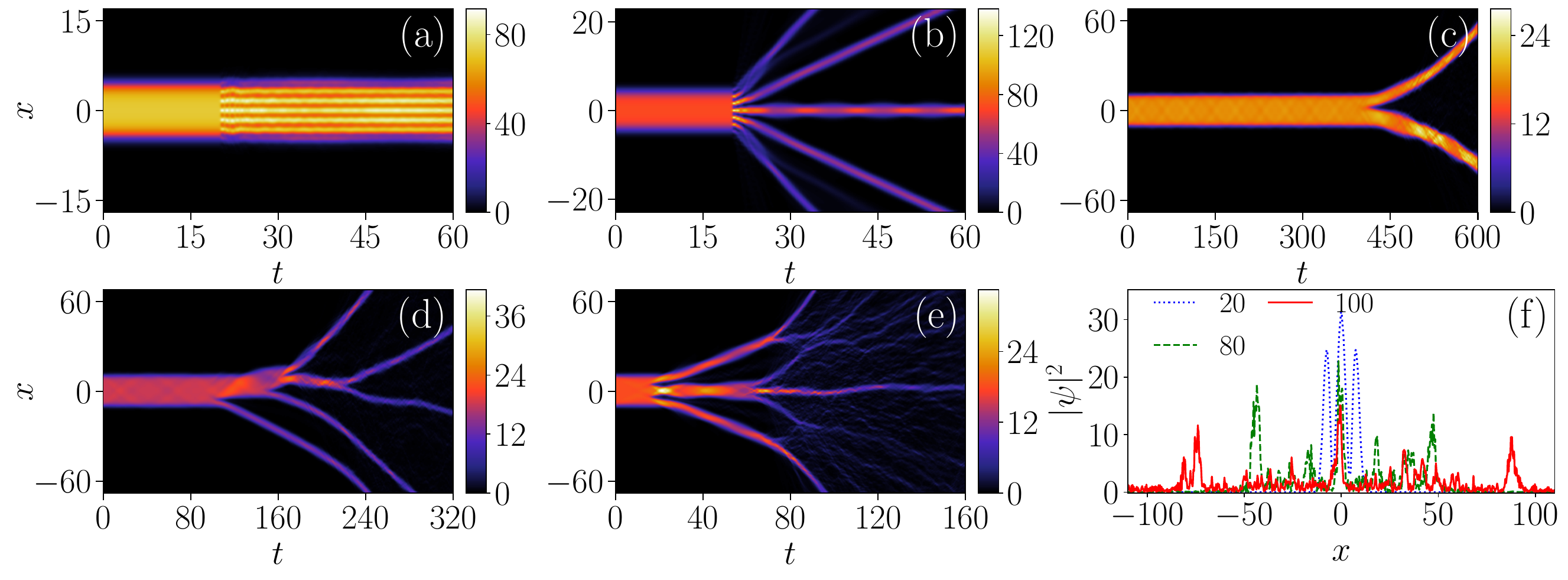}
\caption{Dynamics of the QD following quenches of the Rabi-coupling $\Omega$. Density evolution of the QD after a sudden change of $\Omega=0$ at $t=20$ to (a) $\Omega_f=1$ and (b) $\Omega_f=5$. The initial ground state is characterized by $g=1$, $g_{\uparrow\downarrow}=-0.95$ ($\delta g/g = 0.05$), $\Omega =0, k_L = 2$ and  {$N=300$}. The quench triggers either the dynamical generation of a quantum stripe droplet [panel (a)] or a fragmented droplet configuration [panel (b)]. Time-evolution of the quenched droplet density starting from $\Omega =1$ to (c)  {$\Omega_f=4$}, (d)  {$\Omega=11$}, and (e)  {$\Omega=30$}. (f) Characteristic density profiles at different time instants (see legend) of a panel (e). The remaining parameters are $g=1$, $g_{\uparrow\downarrow} = -0.9$ ($\delta g/g = 0.10$), $k_L = 0.5$ and  {$N=180$}. The quench gives rise to an unstable evolution associated with multiple droplet-breaking events into asymmetrically moving QDs. Emitted radiation accompanies the dynamics, especially for larger $\Omega_f$.}
\label{fig:quen3}
\end{figure*}

The resulting dynamics of the quantum  {droplet} (QD) total density after applying a quench of the Rabi-coupling from $\Omega=0$ to distinct finite values are demonstrated in Fig.~\ref{fig:quen3}. The QD is initially in its ground state, where $g=1$, $g_{\uparrow\downarrow}=-0.95$ ($\delta g/g = 0.05$), $\Omega =0$, $k_L = 2$, and  {$N=300$}. We note in passing that quenches in $\Omega$ for smaller $N$ mainly result in breathing-like oscillations of the droplet and will not be discussed below. The droplet is left to freely evolve for a finite time, and subsequently, at $t\sim 20$, $\Omega$ is quenched. As expected, until $t=20$, the QD maintains its shape [Fig.~\ref{fig:quen3}(a), (b)] since it corresponds to the unperturbed ground state of the system. However, around $t \sim 20$, the QD structurally deforms due to the modified $\Omega$, and, in particular, stripes build upon the density background resembling a stripe QD.
In the case of a post-quench $\Omega_{{\rm f}}=1$, the emergent striped pattern appears to be robust in the course of the evolution [Fig.~\ref{fig:quen3}(a)]. Namely, it retains its overall structure with the individual stripes performing weak amplitude oscillations after their nucleation and later on remaining quasi-stationary. This density pattern is reminiscent of the respective ground state configuration for $\Omega=1$ and $k_L=2$. For completeness, the response of stripe droplets to $\Omega$-quenches is discussed in Appendix \ref{stripe_quench}.
In sharp contrast, a larger Rabi-coupling, e.g., $\Omega_{{\rm f}}=5$, leads to a more complex evolution in terms of the fate of the generated configurations. Indeed, after the quench, we observe a fragmentation of the QD into several counterpropagating ones outwards and a central one, see Fig.~\ref{fig:quen3}(b). It is worth noting that the outer droplets appear to accelerate over time (see the bending behaviour of each density branch), and the inner one moves at an almost constant velocity. Additionally, as time progresses, most outer droplets merge, and the central one maintains a constant velocity accompanied by weak amplitude breathing-like oscillations. This final configuration is completely different from what is expected at the ground state level of the system and appears to possess lower energy than the ground state. Specifically, the breaking into multiple droplets is attributed to the energetically unstable nature of the ground state for large values of $\Omega$, which is otherwise dynamically stable, as can be inferred from the spectral analysis of Sec.~\ref{sec:3b}. A similar fragmentation behaviour was observed in binary short-range droplets in \cite{Mithun2020}, triggered by the modulational instability phenomenon.

On the other hand, the quench-induced droplet response changes drastically for post-quench Rabi couplings satisfying $\Omega_{{\rm f}}>k_L^2$. This is demonstrated below in the general case of an initially finite Rabi-coupling $\Omega \neq 0$. Fig.~\ref{fig:quen3}(c)-(e) depicts the time-evolved QD densities regarding quenches at $t=0$ from $\Omega=1$ to higher values, e.g.,  {$\Omega_f=4, 11, 30$}. Here, the initial state is characterized by $\Omega=1$, $k_L=0.5$, and  {$N=180$}, while the interactions are $g=1$ and $g_{\uparrow\downarrow}=-0.9$ ($\delta g/g = 0.10$). Interestingly, it is found that this quenching process gives rise to a quite different structural deformation of the original QD compared to the previous case. This is attributed to the dominant Rabi energy term, which is more negative than the previous quench scenario from $\Omega=0$ to a finite $\Omega_{{\rm f}}$. At the initial stages of the evolution, the QD features internal excitations which, after a finite amount of time, lead to the breaking of the droplet into multiple ones whose arrangement and time of creation depend crucially on $\Omega_{{\rm f}}$. For instance, at  {$\Omega_f=4$}, the underlying instability gets activated around  {$t\sim 450$}, triggering the breaking of the droplet into two outgoing ones possessing equal and opposite velocities, which keep on increasing with time as depicted in Fig.~\ref{fig:quen3}(c). On the contrary, a relatively larger post-quench Rabi-coupling, such as  {$\Omega_f=11$}, is associated with three asymmetric breaking events with respect to $x=0$, and an eventual erratic dynamical droplet arrangement, see Fig.~\ref{fig:quen3}(d). Indeed, the first splitting occurs at  {$t\sim 120$}, where a heavy and slow-moving droplet emerges, travelling to the left, and a lighter one with arguably much larger velocity accelerates towards the right edge. Note the earlier time of the fragmentation process than the one for  {$\Omega_f=4$} [Fig.~\ref{fig:quen3}(d)]. Afterwards, this latter heavy droplet suffers a subsequent fragmentation around  {$t\sim 180$}, where left and right-moving daughter droplets arise. Another splitting event generating two droplets takes place at  {$t\sim 250$}, and this splitting process continues for larger evolution times.

Along the same lines quenches, see Fig.~\ref{fig:quen3}(e), to even larger  {$\Omega_f=30$} result in a similar fragmentation procedure occurring at faster timescales, e.g. at  {$t\sim 20$} where the original droplet breaks into three droplets. Namely, the two outer droplets accelerate towards the left and right edges, respectively, and a central droplet with zero velocity. Here, a non-negligible density emission accompanies this process. Moreover, the droplet at the center undergoes at later times  {$t\sim 80$} a further splitting where again  {three} highly localized moving droplets appear asymmetrically while simultaneously emitting additional density small density portions. The emitted density parts accumulate during the dynamics and, eventually, at longer timescales, the spatial distribution becomes highly delocalized. The above-described density delocalization caused by an instability of the droplet upon quenching $\Omega$ is better visualized in the corresponding density snapshots provided in Fig.~\ref{fig:quen3}(f) for  {$\Omega_f=30$}. It is important to mention at this point that the ground state of the SOC droplet referring to  {$\Omega=4, 11, 30$} should be dynamically stable as discussed via the eigenspectrum in Sec.~\ref{sec:3b}. However, the energy of the ground state for  {$\Omega=4, 11, 30$} is higher than that of the post-quench state during the evolution, which is probably an admixture of excited states. For this reason, the source of the observed instability here can be attributed to the presence of an energetically unstable ground state~\cite{Ozawa2013}.

\subsection{Response to SOC wavenumber modifications}
\label{sec:3f}

\begin{figure*}[!htp]
\centering\includegraphics[width=0.99\linewidth]{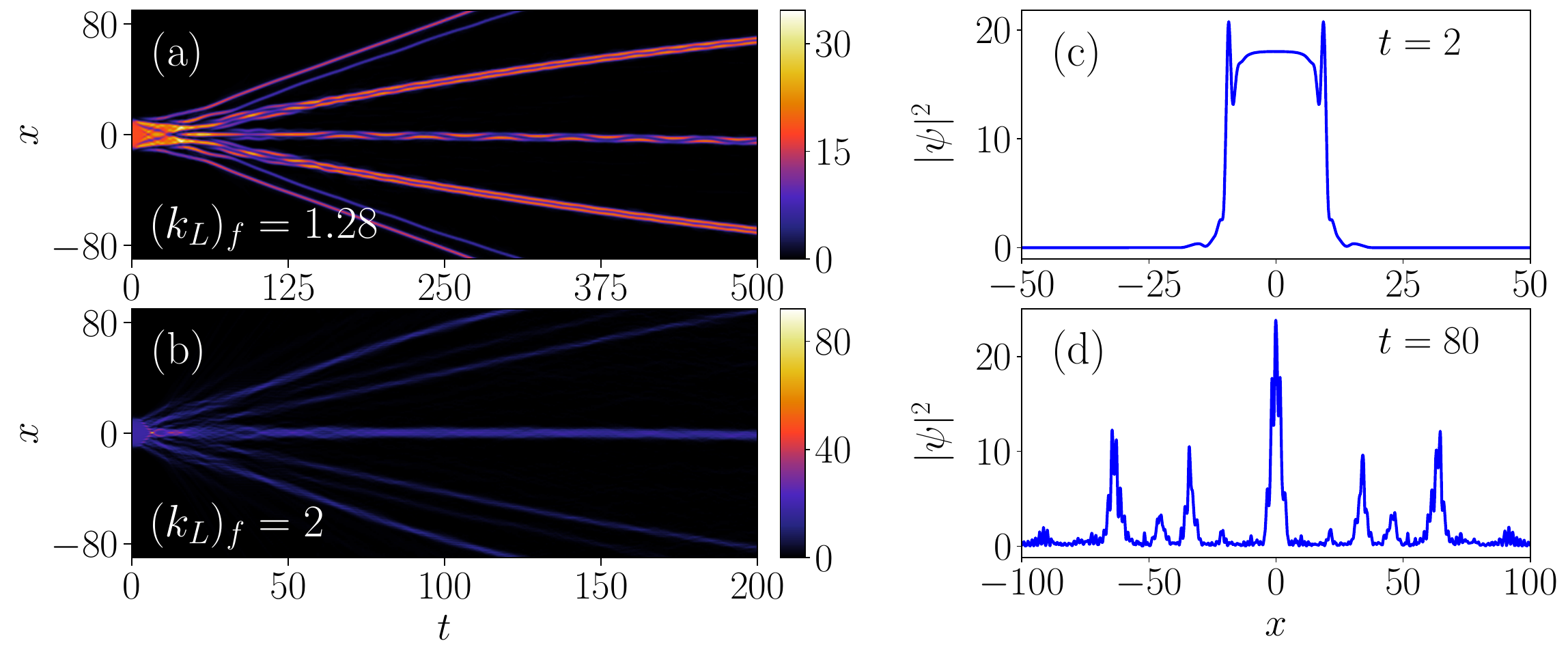}
\caption{Total density evolution of the SOC droplet upon quenching the SOC wavenumber from $k_L=0.5$ to (a)  {$(k_L)_f=1.28$} and (b) $(k_L)_f=2$. 
(c), (d) Profiles of the total density of panel (b) at specific time-instants (see legends). Depending on the quench amplitude, the droplet fragments either into an admixture of counterpropagating single and stripe droplets [panel (a)] or solitary wave patterns exhibiting an overall strong delocalization [panel (b)]. The interactions of the droplet are $g=1$ and $g_{\uparrow\downarrow}=-0.9$ ($\delta g/g = 0.1$), while the Rabi coupling is fixed to $\Omega =1$ and  {$N=180$}.}
\label{fig:quen6}
\end{figure*}

Finally, we analyze the droplet dynamics after quenching the SOC wavenumber, $k_L$, as shown in Fig.~\ref{fig:quen6}. Assuming $g=1$, $\delta g/g = 0.1$, $\Omega =1$,  {$N=180$}, and an initial $k_L=0.5$, the droplet resides in the FT region with no spatial undulations. In this case, the unstable droplet dynamics manifest differently compared to the $\Omega$-quench shown in Fig.~\ref{fig:quen3}(a) and (b). Focusing on the quench to  {$k_L=1.28$} [Fig.~\ref{fig:quen6}(a)], we observe that the original droplet breaks into distinct ones already at short timescales $t\sim 20$. This is a consequence of the prevailing contribution of the SOC attractive term. Namely, two outer droplets move towards the left and right edges, two travel with opposite velocities towards $x=0$, and a heavier one remains at the center, shrinking in size. During this process, further emission events from the outer droplets occur, and later on, a collision of the central droplet with its neighbouring droplets is observed around  {$t \sim 40$}. This collision subsequently leads to the generation of a multitude of stripe droplets, i.e. density branches containing stripes. Soon after their nucleation, these stripe droplet patterns feature internal collisions, due to their excited nature, which give rise to mass transfer phenomena among the individual stripes. After this collisional stage  {($t>150$)}, the individual droplet branches possess a definite shape. Specifically, there are two outer counterpropagating droplet pairs, each with a single  {stripe} droplet and almost equal velocities, two slower oppositely moving droplet stripe branches, and a central single droplet that remains more or less stationary. All the above-mentioned droplet fragments feature breathing-like excitation. This latter phenomenon results in mass exchange among the droplets within the slow-moving stripe branches.

Increasing the quench amplitude, e.g. to $(k_L)_f = 2$, is associated with a faster breaking (around  {$t \sim 5$}) of the initial droplet and a more prominent eventual delocalization of the droplet density, as shown in Fig.~\ref{fig:quen6}(b). In contrast to the previous scenario, stripe droplet branches do not form, which is attributed to the larger quench amplitude causing more excitations into the system. The central droplet appears to be the most excited and experiences a pronounced breathing behaviour. Here, the long-time localized density branches show solitary wave characteristics in shape, as we have confirmed by fitting them to bright soliton solutions~\cite{kevrekidis2008emergent}. To demonstrate the spatial delocalization of the droplet distribution, Figs. \ref{fig:quen6}(c) and (d) illustrate corresponding density profiles at different instants. It becomes clear that initially, the unstable behaviour develops near the surface of the droplet [Fig.~\ref{fig:quen6}(c)], and over time, it spreads in space leading to several highly localized symmetric density branches with respect to $x=0$, featuring a delocalized background due to multiple emissions (radiation), see Fig.~\ref{fig:quen6}(d). 
 {The above-discussed unstable behavior of the SOC droplet 
following the quench in $k_L$ can be attributed, at least partially, to the dynamically unstable phase (for $\Omega < k_L^2$) of the droplet in the final state as suggested by 
the excitation spectrum presented in  Fig.~\ref{fig:bdg-spectrum1}. 
We also note the absence of dynamical instability for quenches at which 
the final state satisfies $\Omega > k_L^2$. 
This is in line with the observations made in the respective excitation spectrum.} 

\section{Summary and Perspectives}
\label{sec:4}

We have studied the stationary properties and nonequilibrium dynamics of droplet configurations appearing in SOC 1D bosonic mixture. To describe these many-body bound states of matter, we rely on the eGPE approach that accounts for the lowest-order quantum LHY correction. The stationary configurations are investigated for varying mean-field interactions, number of atoms, and SOC parameters. It is explicated that for small SOC wavenumbers, a transition from a Gaussian to an FT structure takes place for either increasing atom number or decreasing intercomponent attraction with all other parameters fixed. However, using a larger SOC wavenumber and finite interactions, we showcase the appearance of stripe patterns upon the droplet distribution and the eventual saturation of the background density to an FT. To further understand the aforementioned transition behaviour, e.g. in terms of the atom number, we calculate the system's chemical potential and kinetic energy. The latter attains its maximum value at the transition threshold, with the respective particle number being smaller for decreasing the magnitude of intercomponent attraction. Furthermore, we estimate the surface energy of the droplet, which is found to obey a power-law dependence on the relative interaction ratio, and it is the same independent of the stripe character of the droplet. 

Importantly, by exploiting the excitation spectrum of the above-described droplet configurations, we can infer their stability properties. 
{For instance, at small SOC wavenumbers, Gaussian-shaped and FT droplets are stable in terms of different parametric variations of the intercomponent attractive interaction and the atom number. 
In sharp contrast, for larger SOC wavenumbers, we find that stripe droplets are unstable configurations as dictated by the emergence of complex eigenvalues in the excitation spectrum independently of the particle number or intercomponent attraction.}

The droplet dynamics are analyzed for different quench protocols, namely: (i) imprinting an initial velocity, (ii) utilizing intercomponent interaction quenches, and (iii) considering sudden changes of either the SOC wavenumber or the Rabi coupling. The emergent patterns are attributed to the prevailing nature of specific energy contributions in each case. For velocity perturbations, we find that the droplet undergoes breathing-like oscillations for small post-quench velocities or breaks into several moving droplets for larger amplitude quenches. This phenomenology occurs independently of the SOC wavenumber, i.e., the striped nature of the droplet structure, and it can be traced back to the dominant character of the attractive SOC energy term. The critical velocity for the transition from the breathing to the moving droplet is maximal around the particle number threshold at which the ground state transits from a Gaussian to FT. A similar trend is also evident in the breathing frequency of the droplet, which is generally smaller for stripe droplets. It increases with the atom number reaching a maximum at the transition from Gaussian to FT droplets and afterwards satisfies a power-law decrease inversely proportional to the particle number. The latter signals the phononic spectrum of FT droplets, irrespective of the SOC wavenumber.

The droplet response following intercomponent interaction quenches is also diverse. In the case of reduced attraction, the droplet either expands for larger quench amplitudes or splits into two counterpropagating fragments. However, performing quenches to stronger attractions results in a breathing droplet, and for larger amplitudes, the droplet breaks into several fragments whose number increases with larger attraction. Employing quenches of the Rabi-coupling leads to a structural deformation from a smooth droplet to a striped one for relatively small quench amplitudes. Additionally, asymmetric droplet fragmentation occurs for increasing post-quench Rabi-coupling, where the latter exceeds the SOC (spin-orbit coupling). Interestingly, abrupt changes of the SOC wavenumber again allow the breaking of the original droplet due to the dominance of the attractive SOC contribution. This can lead to admixtures of single and striped droplet branches or erratic spatial distributions towards larger wavenumbers.

There are several interesting future research directions that can be pursued based on our results. It would be particularly relevant to investigate the correlation properties~\cite{mistakidis2023few} of the discussed stripe droplet configurations and engineer anti-ferromagnetic two-body configurations. On the other hand, considering the same 1D geometry, it would be worth identifying the different families of solutions that exist for relevant parametric variations, including the SOC wavenumber and Rabi coupling. Another extension is to employ the generic two-component setup by breaking the underlying SU(2) symmetry to reveal the different phases that can arise bearing miscible, immiscible and more complex arrangements of the spin components. Certainly, the generalization of our analysis to two dimensions aiming to understand the excitation spectrum of the stripe patterns and their relation with supersolids~\cite{chomaz2022dipolar} is an intriguing perspective. 

\acknowledgements
S.G. would like to acknowledge the financial support from the University Grants Commission - Council of Scientific and Industrial Research (UGC-CSIR), India. We thank Rony Boral for the discussion. We also gratefully acknowledge our supercomputing facilities Param-Ishan and Param-Kamrupa (IITG), where all the numerical simulations were performed. R.R. acknowledges the postdoctoral fellowship supported by Zhejiang Normal University, China, under Grants No. YS304023964. S.I.M.~acknowledges support from the NSF through a grant for ITAMP at Harvard University. P.M. acknowledges the financial support from MoE RUSA 2.0 (Bharathidasan University - Physical Sciences).

\appendix

\counterwithin{figure}{section}


\section{Matrix elements of the Bogoliubov-de-Gennes equation}\label{app:bdg}
\begin{figure*}[!htp]
\centering\includegraphics[width=0.98\linewidth]{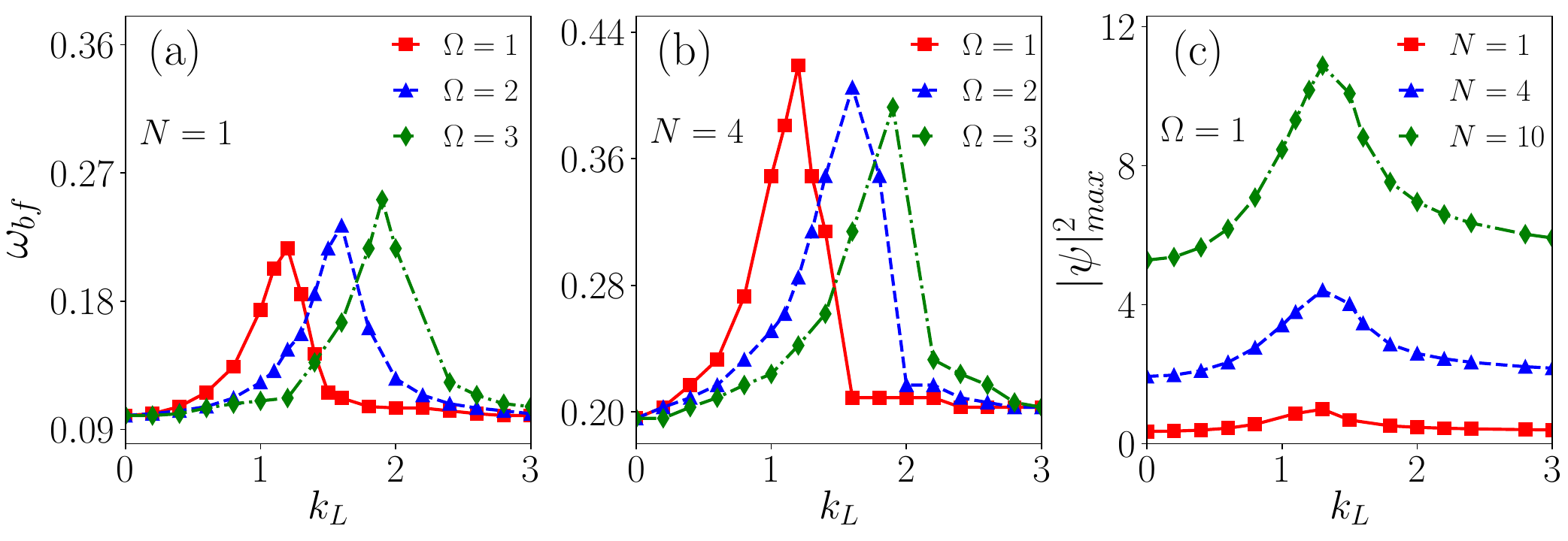} 
\caption{Breathing frequency, $\omega_{bf}$, with respect to the SOC wavenumber $k_L$ for various $\Omega$ (see legends) at (a) $N=1$ and (b) $N=4$. The breathing motion is generated by perturbing the droplet with velocity $v=0.05$ with the ground state prepared with $\delta g/g=0.1$. For all $N$, $\omega_{bf}$ first increases for larger $k_L$ below a certain threshold $k^c_L$ and beyond which it decreases. The maximum $\omega_{\mathrm{bf}}$ is larger for increased $\Omega$ when $N=1$, while it shows a decreasing behaviour for $N=4$. (c) The peak density, $\vert \psi\vert ^2_{max}$, for different $k_L$ and several fixed $N$ (see legend) for fixed $\Omega=1$. The peak density increases with $N$ and acquires a maximum for all $N$ at the transition where the SOC wavenumber is $k^c_L$.}
\label{fig:brefre2}
\end{figure*}

 {In this appendix, we provide the explicit matrix elements of the eigenvalue problem used in Section~\ref{sec:3b} to obtain the excitation spectrum of the SOC droplet solutions. The case of equal intracomponent interactions, $g_{\uparrow \uparrow}=g_{\downarrow \downarrow}\equiv g$, is considered. The individual matrix elements of Eq.~(\ref{eq:bdgEigenval}) exploiting the symmetries of the system read %
\begin{align}
Y_1 & = \left[\dfrac{\delta g}{2}+g -\alpha \right]\psi_{\uparrow}^2, \;\;
Y_2 = \left[\dfrac{\delta g}{2}+g -\alpha \right]\psi_{\downarrow}^2\\
Z_1 & =\left[\dfrac{\delta g}{2}-g-\alpha\right]\psi_{\uparrow}\psi^*_{\downarrow}-\Omega , \;\;
Z_2 = \left[\dfrac{\delta g}{2}-g -\alpha\right]\psi_{\uparrow}\psi_{\downarrow}
\end{align}
and %
\begin{align}
f_1 = & -\frac{1}{2} \partial_x^2 - \mathrm{i} k_L \partial_x + (2 g + \delta g - 3 \alpha) \lvert\psi_{\uparrow}\rvert^2 \notag \\ 
& + \left( \frac{\delta g}{2}-g - 2 \alpha \right) \lvert \psi_{\downarrow}\rvert^2 - \mu_{\uparrow} \\
f_2 = & -\frac{1}{2} \partial_x^2 + \mathrm{i} k_L \partial_x + \left( \frac{\delta g}{2}-g - 2 \alpha \right) \lvert\psi_{\uparrow}\rvert^2 \notag \\ 
&+ (2 g + \delta g - 3 \alpha) \lvert\psi_{\downarrow}\rvert^2 - \mu_{\downarrow},
\end{align}
where $\alpha = g_{\mathrm{LHY}}^{3/2}/\left(2\pi \sqrt{\lvert\psi_{\uparrow}\rvert^2+\lvert\psi_{\downarrow}\rvert^2}\right)$. 
The chemical potentials of the individual components assume the forms~\cite{Gangwar2022}
\begin{subequations}
\label{eq:gpsoc2:mu:ab}
\begin{align}
 \mu_{\uparrow} = 
 & \frac{1}{N_\uparrow} \int \Bigg[\frac{1}{2}\left\vert \frac{\partial \psi_{\uparrow}}{\partial x}\right\vert^2 + \tilde H_\uparrow + \left( k_{L} \frac{\partial \psi^I_{\uparrow}}{\partial x} + \Omega \psi^R_{\downarrow} \right) \psi^R_{\uparrow } \notag \\ 
 & - \left( k_{L} \frac{\partial \psi^R_{\uparrow R}}{\partial x} - \Omega \psi^I_{\downarrow} \right) \psi^I_{\uparrow} \Bigg] dx , \label{eq:gpsoc2:mu:1} \\
 \mu_{\downarrow} = 
 & \frac{1}{N_\downarrow} \int \Bigg[ \frac{1}{2} \left\vert\frac{\partial \psi_{\downarrow }}{\partial x}\right\vert^2 + \tilde H_\downarrow - \left( k_{L} \frac{\partial \psi^I_{\downarrow}}{\partial x} - \Omega \psi^R_{\uparrow} \right) \psi^R_{\downarrow} \notag \\ 
 & + \left( k_{L} \frac{\partial \psi^R_{\downarrow}}{\partial x} + \Omega \psi^I_{\uparrow} \right) \psi^I_{\downarrow} \Bigg] dx , \label{eq:gpsoc2:mu:2}
\end{align}
where
\begin{align}
N_j = & \int \lvert\psi_j\rvert^2 dx, \;\; j = \uparrow, \downarrow, \\
\tilde H_\uparrow = & \left[g \lvert \psi_\uparrow \rvert^2 
+ g_{\uparrow\downarrow} \lvert \psi_\downarrow \rvert^2 +\dfrac{\delta g}{2}\left(\lvert \psi_{\uparrow}\rvert^2-\lvert\psi_{\downarrow}\rvert^2\right) \right.\notag \\ & \left. - \frac{g^{3/2}}{\pi} \sqrt{\lvert \psi_{\uparrow} \rvert^2 + \lvert \psi_{\downarrow}\rvert^2} \right] \lvert\psi_{\uparrow } \rvert^2 ,\\
\tilde H_\downarrow = & \left[ g \lvert \psi_\downarrow \rvert^2+ g_{\downarrow\uparrow} \lvert \psi_\uparrow \rvert^2 +\dfrac{\delta g}{2}\left(\lvert \psi_{\downarrow}\rvert^2 -\lvert \psi_{\uparrow} \rvert^2\right) \right. \notag \\ & \left.- \frac{g^{3/2}}{\pi} \sqrt{\lvert \psi_{\uparrow}\rvert^2+\lvert \psi_{\downarrow}\rvert^2} \right]\lvert\psi_{\downarrow }\rvert^2.
\end{align}
\end{subequations}
In these expressions, the real and imaginary parts of the mixture wave function are denoted by $\psi^R_{\uparrow, \downarrow }$ and $\psi^I_{\uparrow, \downarrow }$, 
respectively. }

\section{Dependence of the droplet breathing frequency on the SOC parameters}\label{app_breath}

Having analyzed the impact of the effective mean-field interaction $\delta g/g$, and atom number $N$ on the breathing frequency $\omega_{bf}$ in the main text, we proceed to examine in more detail the effect of the SOC wavenumber ($k_L$) and Rabi-coupling ($\Omega$). 
Fig.~\ref{fig:brefre2} presents the behaviour of $\omega_{bf}$ as a function of $k_L$ for different Rabi-couplings while maintaining a fixed $N$ in each panel. For small particle number, e.g. $N=1$ shown in Fig.~\ref{fig:brefre2}(a), it is observed that $\omega_{bf}$ increases with $k_L$ until reaching a maximum at a critical $k_L$ value, and then reduces for larger $k_L$. %
\begin{figure}[!ht]
\centering\includegraphics[width=0.99\linewidth]{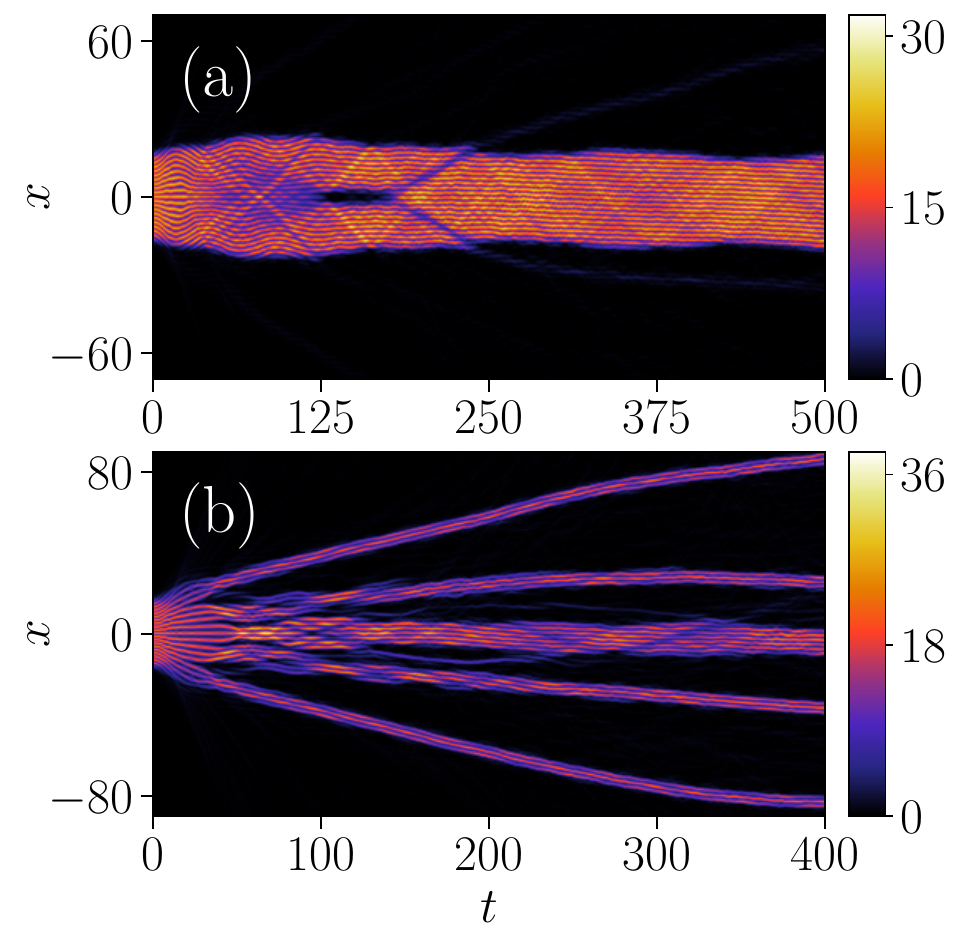}
\caption{Time-evolution of the density of a stripe QD subjected, at $t=0$, to sudden change of the Rabi-coupling from $\Omega =1$ to (a) { $\Omega_f = 2.2$ }and (b) { $\Omega_f =2.8$}. It becomes clear that a weak amplitude quench on $\Omega$ [panel (a)] gives rise to internal excitations as well as droplet breaking and subsequent merging. 
However, larger quench amplitudes [panel (b)] lead to a pronounced breaking into multiple counter propagating stripe droplets showing breathing-like oscillations. 
The remaining system parameters correspond $g=1$, $g_{\uparrow\downarrow}=-0.9$ ($\delta g/g = 0.10$),  {$N=280$} and $k_L = 2$.}
\label{fig:quen9}
\end{figure}%
It eventually approaches the same value as the one at small $k_L$, i.e. $k_L \sim 0$. The increase in the maximum breathing frequency at the transition point can be attributed to the enhanced rigidity resulting from the higher density at that particular point. Moreover, increasing $\Omega$ leads to an elevation in the maximum $\omega_{\mathrm{bf}}$, and the same holds for the critical $k^c_L$ above which $\omega_{\mathrm{bf}}$ starts to decrease. The exact opposite behaviour in terms of the maximum breathing frequency with increasing $\Omega$ is found for larger $N$. For instance, at $N=4$, the maximum $\omega_{\mathrm{bf}}$ at critical $k_L$ exhibits a decreasing tendency with $\Omega$ [Fig.~\ref{fig:brefre2}(b)], and this behaviour persists as $N$ further increased to a higher value, e.g., $N=10$. In order to understand the origin of the maximum breathing frequency at the critical SOC wavenumber, $k^c_L$, for different $N$, we provide the peak density (i.e. $\vert \psi \vert^2_{max}$) with $k_L$ for different $N=1,4$, and 10 by fixing $\Omega=1$. It is found that for a fixed $N$, the total density acquires a maximum, $\vert \psi\vert^2_{max}$, at the critical SOC wavenumber. This feature occurs independently of $N$[See Fig.~\ref{fig:brefre2}(c)].

\section{Generation of stripe droplet fragments}
\label{stripe_quench} 

In the main text, Sec.~\ref{sec:3e}, we discussed the dynamical response of a standard (non-modulated) droplet following a quench of the Rabi-coupling term. Here, for reasons of completeness, we aim to briefly analyze the effect of the abrupt change in $\Omega$ when the original (at $t=0$) configuration is a stripe droplet. As such, a SOC setup is prepared with $g=1$, $\delta g/g = 0.1$,  {$N=280$}, $\Omega=1$ and $k_L = 2$. Applying a quench to  {$\Omega_f=2.2$}, see Fig.~\ref{fig:quen9}(a), leads to internal, mainly breathing-like, excitations of the stripe droplet. 
As time evolves, these excitations allow the splitting of the entire stripe structure, around  {$t\sim 125$}, into two symmetric ones, which naturally contain a smaller amount of stripes. Later on,  {$t\sim 180$}, they merge into a huge breather-striped droplet, which remains as such during the evolution. This merging process is accompanied by the simultaneous emission of two counter-propagating (with nearly equal velocity) density branches towards the edges. 

On the other hand, the emergent dynamical response appears to be more drastically altered utilizing a quench to larger Rabi-coupling such as  {$\Omega_f=2.8$} as can be seen in Fig.~\ref{fig:quen9}(b). The modification in $\Omega$ causes the internal excitation of the original striped pattern, which after some time evolution, emits density fractions. This latter event initiates collisions between the nearest neighbouring stripes and, afterwards, the breaking of the entire structure into multiple counter-propagating fragments consisting of stripe droplets. These striped density branches suffer breathing motion and mass transfer among themselves, but they survive during evolution. 


\bibliography{references.bib}

\end{document}